\documentclass[12pt]{article}      

\title{Quantum Calculus of Fibonacci Divisors and Infinite Hierarchy of Bosonic-Fermionic Golden Quantum Oscillators}

\author{ Oktay K Pashaev \\Department of Mathematics\\ Izmir Institute of Technology \\ Urla-Izmir, 35430, Turkey}

\begin{document}
\newcommand{\be}{\begin{equation}}
\newcommand{\ee}{\end{equation}}
\newcommand{\bea}{\begin{eqnarray}}
\newcommand{\eea}{\end{eqnarray}}
\newcommand{\disp}{\displaystyle}
\newcommand{\la}{\langle}
\newcommand{\ra}{\rangle}

\newtheorem{thm}{Theorem}[subsection]
\newtheorem{cor}[thm]{Corollary}
\newtheorem{lem}[thm]{Lemma}
\newtheorem{prop}[thm]{Proposition}
\newtheorem{definition}[thm]{Definition}
\newtheorem{rem}[thm]{Remark}
\newtheorem{prf}[thm]{Proof}

\maketitle


\begin{abstract}
Starting from divisibility problem for Fibonacci numbers we
introduce Fibonacci divisors, related hierarchy of Golden derivatives in powers of the Golden Ratio and develop corresponding quantum calculus.
By this calculus, the infinite hierarchy of Golden quantum oscillators with integer spectrum determined by Fibonacci divisors,
the hierarchy of Golden coherent states and related Fock-Bargman representations  in space of complex analytic functions are derived.
 It is shown that
Fibonacci divisors with even and odd $k$ describe Golden deformed bosonic and fermionic quantum oscillators, correspondingly.
By the set of translation operators we find the hierarchy of Golden binomials and related  Golden analytic functions,  conjugate to Fibonacci number $F_k$.
In the limit $k \rightarrow 0$,  Golden analytic functions reduce to  classical holomorphic functions
and quantum calculus of Fibonacci divisors  to the usual one.
 Several applications of the calculus to
 quantum deformation of bosonic and fermionic oscillator algebras, R-matrices, hydrodynamic images and quantum computations are discussed.
\end{abstract}


\section{Introduction}

The Golden calculus is the quantum or $q$-calculus with Golden Ratio bases $\varphi$ and $\varphi'$, so that Binet formula for Fibonacci
numbers becomes the $q$-number in this calculus \cite{golden}.
The corresponding Golden derivative appears as a finite q-difference derivative with Golden ratio base,
 allowing one to construct Golden binomials and Taylor expansion in terms
 of these binomials. According to this expansion, the Golden exponential functions  and corresponding trigonometric and
 hyperbolic functions, as solutions of Golden ODE of oscillator type and PDE of wave type have been derived \cite{golden}.

It was shown that quantized Fibonacci number operator determines  the deformed Golden quantum oscillator with discrete energy spectrum in the form of Fibonacci numbers \cite{golden}. The relative difference of the energy levels for this oscillator asymptotically, for big $n$ becomes the Golden ratio.
It turns out that Golden calculus as $q$-calculus with fixed bases $\varphi, \varphi'$ does not allow  limit $q \rightarrow 1$  to the usual calculus. In the present
paper we address this problem by working with Fibonacci divisors, instead of Fibonacci numbers.
In general, the ratio of two Fibonacci numbers  is not an integer number. However, surprising fact is that Fibonacci numbers of the form $F_{k n}$, where $k$ and $n$ are arbitrary integers,
is dividable by Fibonacci number $F_k$. The infinite sequence of integer numbers  $F_{k n}/F_k \equiv F^{(k)}_n$ we call the
Fibonacci divisors conjugate to $F_k$ or shortly, the $k-th$ Fibonacci divisors. According to number $k$ we have an infinite hierarchy of Fibonacci divisor sequences.  An intriguing fact is that the Binet type formula for $F^{(k)}_n$ divisor  is determined by integer $k$-th power of the Golden ratio $\varphi$ and $\varphi' = -1/\varphi$,
\be
F^{(k)}_n = \frac{(\varphi^k)^n - ({\varphi'}^k)^n}{\varphi^k - {\varphi'}^k}.\label{Fnk}
\ee
 For $k=1$ this formula gives the sequence of usual Fibonacci numbers, while for $k=2,3,4,...$ we have generalized Fibonacci sequences, with three term recurrence relations, depending on value of $k$. The goal   of the present paper is to develop quantum calculus for these hierarchy of number sequences and construct an infinite hierarchy of Golden oscillators, with discrete spectrum determined by these numbers.

The paper is organized as follows. In Section 2 we discuss briefly some physical examples from quantum mechanics, hydrodynamics, quantum information theory and
quantum integrable systems, where  Fibonacci divisors $F_n^{(k)}$ and corresponding calculus can be applied.
Division of Fibonacci numbers and algebraic properties of corresponding integers, denoted as Fibonacci divisors are subject of Section 3.
In Section 4 we develop corresponding calculus and introduce the hierarchy of Golden derivatives, corresponding $k-th$ Golden periodic functions  and generating function for
Fibonacci divisors. Section 5 deals with entire generating function for Fibonacci divisors and several interesting identities for them.
The $k-th $ hierarchy of Fibonomials, constructed from $F_n^{(k)}$ is subject of Section 6. The corresponding hierarchy of Golden binomials, the $k-th$ Golden binomial expansion and
Golden Taylor formula are developed in  Section 7. By this formula the set of exponential functions, the translation operator and $k-th$ hierarchy of Golden analytic functions are introduced. In Section 8 we apply calculus of Fibonacci divisors  to hierarchy of bosonic and fermionic Golden quantum oscillators. Finally, the hierarchy of Golden coherent states and Fock-Bargman representations are obtained as realization of quantum calculus of Fibonacci divisors in complex plane.

\section{Physical Motivations of $F_n^{(k)}$}

Below we  describe several physical systems in which $F^{(k)}_n$  appears naturally.

\subsection{$F_N^{(k)}$ and the Golden-deformed bosons and fermions}

For even $k$ equation (\ref{Fnk})  in the limit $k \rightarrow 0$ gives just natural numbers, $\lim_{k \rightarrow 0} F^{(k)}_n = n$.
In contrast, for odd $k$ in this limit we have the binary numbers 0 and 1, for even and odd $n$ correspondingly. This shows that depending on $k$,
Fibonacci divisor numbers (\ref{Fnk}) can describe deformed bosonic and fermionic quantum oscillators.
These oscillators are determined by  the hierarchy of quantum deformed algebras

\be  b_k b_k^+ - {\varphi'}^k b_k^+ b_k = \varphi^{k N},
\ee
parameterized by integer $k$, with number operator
\be
b^+_k b_k =
F^{(k)}_N = \frac{(\varphi^k)^N - (\varphi'^k)^N}{\varphi^k - \varphi'^k}.\label{FNk0}
\ee
Depending on $k$ we have two type of models.

1) For odd $k = 2l+1$ the algebra
\be  b_k b_k^+ + \frac{1}{\varphi^k}\, b_k^+ b_k = \varphi^{k N},\label{odd}
\ee
becomes a non-trivial $q$-deformed fermionic algebra \cite{PV1},\cite{PV2} for the $q$-deformed quantum oscillators. It was  applied to several problems, as the dynamic mass generation of quarks and nuclear pairing \cite{TL1}, \cite{TL2}, and as descriptive of higher
order effects in many-body interactions in nuclei \cite{sviratcheva}, \cite{ballesteros}.
 When $0 < q < 1$, an arbitrary number of $q$-fermions in this algebra  can occupy a given state.
 If we denote  $q = \frac{1}{\varphi^{2k}}$, where $k$ is positive odd number, the inequality is satisfied and
 algebra (\ref{odd}) coincides with the one in \cite{PV1}.
   The Fock space construction for the Golden-deformed fermionic algebra requires to introduce the "fermionic $q$-numbers" \cite{PV1},
  in the form  (\ref{Fnk}),  for odd $k$,
   \be [n]^F_{\frac{1}{\varphi^{2k}}} = \frac{\varphi^{k n} - (-1)^n \varphi^{-kn}}{\varphi^k + \varphi^{-k}} = F^{(k)}_n \label{fqnumber}.\ee
It shows that for odd $k$ the Fibonacci divisors number operator (\ref{FNk0}) is a specific realization of  fermionic q-number operator of Parthasarathy and Viswanathan \cite{PV1}.
Statistical properties of corresponding  q-deformed fermions, as descriptive of fractional statistics were investigated in \cite{Chaichian}.
As was shown \cite{Narayana},  thermodynamics of these generalized fermions should involve the $q$-calculus with
$q$-derivative in the form
\be D_x \, f(x) = \frac{1}{x} \frac{f(q^{-1} x) - f(-q x)}{q + q^{-1}},\ee
which for
$q = \frac{1}{\varphi}$, becomes the Golden Derivative \cite{golden}. In a more general case $q = \frac{1}{\varphi^k}$, for odd $k$ it becomes
the $k$-th Golden derivative (\ref{hoGoldenDerivative}), studied in the present paper.

2)  For even $k = 2l$ the algebra
\be  b_k b_k^+ - \frac{1}{\varphi^k}\, b_k^+ b_k = \varphi^{k N},\label{even}
\ee
becomes the $q$-deformed bosonic algebra, with number operator
\be
F^{(k)}_N = \frac{(\varphi^k)^N - (\varphi^{-k})^N}{\varphi^k - \varphi^{-k}} = \frac{\sinh N(k \ln \varphi)}{\sinh (k \ln \varphi)}.
\ee
 This algebra and the number operator correspond to quantum algebra with symmetric $q$-numbers, where $q = \varphi^k$ or in notation of paper \cite{PV1}, $q = \frac{1}{\varphi^{2k}}$.

The above results indicate on intriguing relation between divisibility of Fibonacci numbers for odd and even index $k$, and hierarchy of deformed
fermionic and bosonic quantum algebras correspondingly.
 As would be shown below all these numbers are integer, and the spectrum of corresponding oscillators is
also integer.

The Fibonacci divisors number operator,  rewritten in the Fock-Bargman representation   $F_N^{(k)}= F_{z \frac{d}{dz}}^{(k)}$  , becomes the $k$-th
Golden derivative dilatation operator
 \bea
z\, _{(k)} D^{z}_{F}[f(z)]=\frac{(\varphi^k)^{z\frac{d}{dz}} -  ({\varphi'}^k)^{z\frac{z}{dz}} }{\varphi^k -  {\varphi'}^k}f(z)
= \frac{f(\varphi^{k} z)-f(\varphi'^{k} z)}{\left(\varphi^{k} -\varphi'^{k}\right)}, \label{hoGoldenDerivative}
\eea
acting on holomorphic wave function $f(z)$. In this form it can be related with the method of images in hydrodynamics, described in next section.

\subsection{Hydrodynamic images and $k$-th Golden derivatives}

  As was proposed in \cite{Pashaev2019},  an arbitrary  complex analytic function describing quantum state in Fock-Bargman representation, can be interpreted as the complex potential of incompressible and irrotational hydrodynamic flow in two dimensions.   In several bounded domains,  according to
   the method of images, this flow is described by $q$-periodic functions
 and corresponding theorems, such as the two circle theorem \cite{2circle}, the wedge theorem \cite{variation} and the strip theorem \cite{eski}.
 Depending on geometry of the domain, parameter $q$ has different geometrical meanings and values.
For the two circle theorem it is given by squared ratio of two circle radiuses   $q = r^2_2/r^2_1$.
By choosing annular domain between  concentric circles with radiuses in the Golden ratio, we able to give then hydrodynamic interpretation of our $k$-th Golden derivatives and corresponding periodic functions. For even and odd $k$ results are different and would be considered separately.

1) For even $k = 2l$,  derivative  (\ref{hoGoldenDerivative}) is determined by finite difference
\bea
z\, _{(k)} D^{z}_{F}[f(z)]= \frac{f(\varphi^{k} z)-f(\frac{1}{\varphi^{k}} z)}{\left(\varphi^{k} - \frac{1}{\varphi^{k}}\right)} \label{EvenhoGoldenDerivative}
\eea
between values of function $f$ at points $\varphi^k z$ and $\frac{1}{\varphi^{k}} z$. Geometrically, numbers $\varphi^k $ and $\frac{1}{\varphi^{k}} $
are symmetric points for unit circle at origin. The analytic function $f(z)$ describes irrotational and incompressible hydrodynamic flow in two dimensions.
If this flow is in annular domain bounded by two concentric circles $1 < |z| < \sqrt{\varphi}$, then the method of images in the form of two circle theorem
can be applied \cite{2circle}.  Replacing the boundary by an infinite set of images it gives complex potential of the flow in following form
\be
G_\varphi(z) = f_{\varphi}(z) + \bar f_{\varphi}\left( \frac{1}{z}\right),\label{Gphi}
\ee
where
\be
f_{\varphi}(z) = \sum_{n=-\infty}^{\infty} f(\varphi^n z),\,\,\,\,\,\,\,\,\,\bar f_{\varphi}\left(\frac{1}{z}\right) = \sum_{n=-\infty}^{\infty} \bar f\left(\varphi^n \frac{1}{z}\right).
\ee
Here, the first sum describes the image flow in even annular domains and the second sum, in odd ones. Then, the whole complex plane is divided
to infinite set of circles with Golden ratio of successive radiuses.
The complex function $f_{\varphi}(z)$ is the Golden periodic function, $f_{\varphi}(\varphi z) = f_{\varphi}(z)$
and as follows, it valid also for function $G_\varphi(\varphi z) = G_\varphi(z)$. The last relation implies that
\be
\, _{(k)} D^{z}_{F} G_\varphi(z) = 0,
\ee
for any integer $k$. The complex velocity corresponding to  flow (\ref{Gphi}),
\be
\bar V(z) = \frac{d G(z)}{dz} = \sum_{n=-\infty}^{\infty} \varphi^n \bar v(\varphi^n z) - \frac{1}{z^2}\sum_{n=-\infty}^{\infty}\varphi^n v\left(\varphi^n \frac{1}{z}\right),
\ee
is then the Golden self-similar function $\bar V(\varphi z)= \frac{1}{\varphi} \bar V(z)$.

In more general case, the annular domain is bounded by circles $1 < |z| < \varphi^{\frac{k}{2}}$ with fixed $k > 1$, so that the flow is  $k$-th Golden periodic $G_{\varphi^k}(\varphi^k z) = G_{\varphi^k}(z)$, and as follows
\be
\, _{(k)} D^{z}_{F} G_{\varphi^k}(z) = 0.
\ee
 Corresponding  complex velocity  is the  $k$-Golden
self-similar function $\bar V(\varphi^k z)= \frac{1}{\varphi^k} \bar V(z)$, but the flow is not the Golden periodic one $G_{\varphi^k}(\varphi z) \neq G_{\varphi^k}(z)$.

As an example we consider the point vortex of strength $\Gamma$ at position $z_0$ in the annular domain, $1 < |z_0| < \sqrt{\varphi}$, with
complex potential
\cite{2circle}, \cite{PashaevYilmaz},
\be
G(z) = \frac{\Gamma}{2 \pi i}\sum_{n=-\infty}^{\infty} \ln \frac{z - \varphi^n z_0}{z - \varphi^n \frac{1}{\bar z_0}}.\label{G}
\ee
It describes an infinite set of vortex images with Golden ratio $\varphi$ of neighbouring image distances from origin:
for even reflections we have
$$...,\frac{z_0}{\varphi^n} ,...,\frac{z_0}{\varphi^2} ,\, \frac{z_0}{\varphi}, z_0, \,\varphi z_0, \,\varphi^2 z_0,...,\varphi^n z_0,... $$
and for odd reflections
$$...,\frac{1}{\varphi^n \bar z_0},...,\frac{1}{\varphi^2 \bar z_0}, \,\frac{1}{\varphi \bar z_0},\, \frac{1}{\bar z_0},\, \frac{\varphi}{\bar z_0}, \, \frac{\varphi^2}{\bar z_0},..., \frac{\varphi^n}{\bar z_0},... $$
Due to characteristic equation for Golden ratio, $\varphi - 1 = \frac{1}{\varphi}$, the relative distance between two neighbouring images is given by power of Golden ratio,
$|\varphi^{n+1} - \varphi^{n}| = \varphi^{n-1}$ - for even reflections, and $\left|\frac{1}{\varphi^{n}} - \frac{1}{\varphi^{n+1}}\right| = \frac{1}{\varphi^{n+2}}$ - for odd reflections.

Complex potential (\ref{G}) is the Golden periodic one, which means that any vortex image can be chosen as the original vortex, so that the full flow in the plane
will not change by this choice. In addition, following \cite{PashaevYilmaz} the infinite sum for this potential can be  expressed by ratio of first Jacobi elliptic theta functions, with Golden ratio as a parameter.

2) For odd $k = 2l+1$, derivative  (\ref{hoGoldenDerivative}) is given by the finite difference
\bea
z\, _{(k)} D^{z}_{F}[f(z)]= \frac{f(\varphi^{k} z)-f(-\frac{1}{\varphi^{k}} z)}{\left(\varphi^{k} + \frac{1}{\varphi^{k}}\right)} \label{EvenhoGoldenDerivative}
\eea
between values of function $f$ at points $\varphi^k z$ and $-\frac{1}{\varphi^{k}} z$. To interpret these image points geometrically,
we consider the flow in double-circular wedge domain, with boundary restricted by two coordinate lines $\Gamma_1: z=x$ and $\Gamma_2:z= x e^{i\frac{\pi}{2}} = i x$
in first quadrant of complex plane $z$;
and by two concentric circles at origin with radiuses $1$ and $\varphi $; $C_1: z= e^{it}$, and $C_2: z = \sqrt{\varphi}\, e^{it}$, where $0 < t < \frac{\pi}{2}$.
Then, application of  the double-circular wedge theorem \cite{variation} gives complex potential as
\be
G(z) = f_{\varphi}(z) + \bar f_{\varphi}\left( \frac{1}{z}\right),
\ee
where
$$
f_{\varphi}(z) = \sum_{n=-\infty}^{\infty}[ f(\varphi^n z)+f(-\varphi^n z) ],\,\,\,\,\,\,\bar f_{\varphi}\left(\frac{1}{z}\right) = \sum_{n=-\infty}^{\infty}[ \bar f\left(\varphi^n \frac{1}{z}\right)+\bar f\left(-\varphi^n \frac{1}{z}\right)] .
$$
This function is Golden periodic $G(\varphi z) = G(z)$ and in addition, it is even $G(-z) = G(z)$, so that,
\be
_{(k)} D^{z}_{F} G(z)= 0.
\ee
For point vortex in the double-circular wedge domain it gives the sum
\be
G(z) = \frac{\Gamma}{2\pi i} \sum_{n=-\infty}^{\infty} \ln \frac{(z^2 - \varphi^{2n} z^2_0)(z^2 - \varphi^{2(n+1)} \frac{1}{z^2_0})}{(z^2 - \varphi^{2n} \bar z^2_0)(z^2 - \varphi^{2(n+1)} \frac{1}{\bar z^2_0})},
\ee
which is even and arbitrary $k$-th Golden periodic function.
The set of images in this sum includes an infinite number of pairs of  images at antipodal points, $\varphi^n z_0$ and $-\frac{1}{\varphi^n \bar z_0}$,
As we show in next section, these antipodal points play an important role in quantum information theory and can be related to the pair of orthogonal qubit coherent states.

\subsection{$F^{(k)}_n$ and the $n-$qubit coherent states}

The pair of orthogonal qubit states in
coherent state representation
\be
|\psi \ra = \frac{|0\rangle + \psi |1\rangle}{\sqrt{1 + |\psi|^2}} , \,\,\,\,\,\,\,\, |-\frac{1}{\bar\psi} \ra = \frac{-\bar\psi|0\rangle + |1\rangle}{\sqrt{1 + |\psi|^2}},  \ee
is parameterized by complex number $\psi \in C$, given by the stereographic projection $\psi = \tan \frac{\theta}{2} e^{i \phi},$
of the Bloch sphere state $|\theta, \phi \rangle$.
 The states are determined by point $\psi,$ and its antipodal one $- \frac{1}{\bar\psi}$ for unit circle in $C$ \cite{PGNJP}.
Geometrically, they correspond to antipodal points on the Bloch sphere, $M(x, y, z)$ and $M^*(-x, -y, -z)$. For the Golden ratio $\varphi = \tan \frac{\theta}{2}$
the states $|\varphi\rangle$ and $|\varphi'\rangle = | - \frac{1}{\varphi}\rangle$ are orthogonal and correspond to antipodal qubit states
$|\theta, 0 \rangle$ and $|\pi - \theta, \pi\rangle$.

Now we introduce the pair of normalized orthogonal antipodal states
\be
|\varphi^k \ra = \frac{|0\rangle + \varphi^k |1\rangle}{\sqrt{1 + {\varphi^{2k}}}} , \,\,\,\,\,\,\,\, |-\frac{1}{{\varphi}^k} \ra = \frac{- \varphi^k|0\rangle + |1\rangle}{\sqrt{1 + {{\varphi}^{2k}}}}  ,  \label{2normalized}\ee
characterized by integer number $k \in Z$.
 In the limiting case, the states become computational basis states: $| 1 \ra$ and $| 0 \ra$ for $k \rightarrow \infty$,
and  $| 0 \ra$ and $| 1 \ra$  for $k \rightarrow -\infty$.

From these states (without normalization) the set of $n$-qubit states   $| \Psi\ra_k $, characterized by Fibonacci divisors   $F^{(k)}_n$
can be constructed. For odd $k = 2l+1$, we define the state
\bea
| \Psi\ra_k =c_0\frac{(|0\ra + \varphi^k | 1\ra)^n - (|0\ra + {\varphi'}^k | 1\ra)^n}{\varphi^k - {\varphi'}^k },    \eea
which is expanded in computational basis as
\bea
| \Psi\ra_k =  c_0 [
F^{(k)}_1 (|10...0 \ra +  ...|00...1 \ra)
 + F^{(k)}_2 (|110...0 \ra +  ...|00...11 \ra) \,
 ... +  F^{(k)}_n |111...1 \ra].\nonumber\eea
Normalization constant for this state is determined by sum
 \be
 c^{-1}_0 = \sqrt{\sum^n_{s=1}   \left( \begin{array}{c} n \\ s   \end{array}   \right)   (F_s^{(k)})^2}
 \ee
and probabilities of measurement are
\be
 P_l = \frac{(F_l^{(k)})^2}{ \sum^n_{s=1}   \left( \begin{array}{c} n \\ s   \end{array}   \right)   (F_s^{(k)})^2}.
 \ee
 The states are entangled.
 For two qubit state with $n=2$ and arbitrary odd $k$, we have the state
 \be
 | \Psi \ra_k = \frac{|01\ra + |10\ra + L_k |11\ra}{\sqrt{2 + L^2_k}},
 \ee
 where $ L_k = \varphi^k + {\varphi'}^k = F^{(k)}_2 $ - are the Lucas numbers. The level of entanglement in this state is determined by concurrence,
 expressed in terms of Lucas numbers
 as decreasing
 function of  $k$,
 \be
 C_k = \frac{2}{2+ L^2_k}.
 \ee
 The maximal value $C_1 = 2/3$ this concurrence reaches  for $k = \pm 1$.

 From another side, from normalized states (\ref{2normalized}) following \cite{PGNJP}, the set of maximally entangled two qubit states can be derived,

\be
|P_{\pm}\ra = \frac{| \varphi^k\ra   | \varphi^k\ra \pm | -\frac{1}{{\varphi}^k}\ra | -\frac{1}{{\varphi}^k}\ra }{\sqrt{2}},\,\,\,\,\,\,\,\,
|G_{\pm}\ra = \frac{| \varphi^k\ra   | -\frac{1}{{\varphi}^k}\ra \pm | -\frac{1}{{\varphi}^k}\ra | {\varphi}^k\ra }{\sqrt{2}}.
\ee
Two of these states are independent of $k$ and in explicit form are just the Bell states
\be
|P_+\ra = \frac{|00\ra + |11\ra}{\sqrt{2}},\,\,\,\,\,\,\,\,|G_-\ra = \frac{|01\ra - |10\ra}{\sqrt{2}},
\ee
while another pair of states is superposition of the Bell states, depending on Fibonacci and Lucas numbers:

1) for even $k=2l$,
\bea
|P_-\ra = -\frac{\sqrt{5} F_k}{L_k} \,\frac{|00\ra - |11\ra}{\sqrt{2}} + \frac{2}{L_k} \,\frac{|01\ra + |10\ra}{\sqrt{2}},\\
|G_+\ra = -\frac{2}{L_k}\, \frac{|00\ra - |11\ra}{\sqrt{2}} - \frac{\sqrt{5} F_k}{L_k} \,\frac{|01\ra + |10\ra}{\sqrt{2}},
\eea

2) for odd $k = 2l+1$,
\bea
|P_-\ra = -\frac{L_k}{\sqrt{5} F_k} \,\frac{|00\ra - |11\ra}{\sqrt{2}} + \frac{2}{\sqrt{5} F_k} \,\frac{|01\ra + |10\ra}{\sqrt{2}},\\
|G_+\ra = -\frac{2}{\sqrt{5} F_k}\, \frac{|00\ra - |11\ra}{\sqrt{2}} - \frac{L_k}{\sqrt{5} F_k} \,\frac{|01\ra + |10\ra}{\sqrt{2}}.
\eea
By using identities:
$5 F_k^2 + 4 = L_k^2$, for even $k=2l$, and $L_k^2 + 4 = 5 F_k^2$, for odd $k = 2l+1$;
we find that the concurrence for these states is maximal $C =1$ and
in the limit $k \rightarrow \pm\infty$  the states reduce to the second pair of Bell states.

\subsection{ $F_N^{(k)}$ and Hecke characteristic equation for R matrix}

The Fibonacci divisors $F_n^{(k)}$ are connected also with quantum integrable systems approach to the theory of quantum groups,
via solution of the Yang-Baxter equation for the $R$-matrix \cite{Faddeev}. The invertible ${\hat R}$ matrix obeys
a characteristic equation, known as the Hecke condition. For two roots $\varphi^k$ and $-\frac{1}{\varphi^k}$,
\be ({\hat R} - \varphi^k)({\hat R} + \frac{1}{\varphi^k}) = 0 \label{Hecke}\ee it is
\be {\hat R}^2 = B_k {\hat R} + I,\label{Hecke1}\ee
where $B_k = \varphi^k - \frac{1}{\varphi^k}$. For odd $k = 2l+1$ these numbers are the Lucas numbers $B_k = \varphi^k + {\varphi'}^k = L_k$
and equation (\ref{Hecke1}) is characteristic equation for Fibonacci divisors $F_n^{(k)}$, satisfying recursion relation (\ref{maintheorem}),
\be
F_{n+1}^{(k)} = L_k F_n^{(k)} + (-1)^{k-1} F_{n-1}^{(k)}.
\ee
For even $k = 2l$ these numbers can be written by Fibonacci numbers  $B_k = \varphi^k - {\varphi'}^k = F_k \sqrt{5}$.

By studying representations of the braid group, satisfying this quadratic relation
a polynomial invariant in two variables for
oriented links was obtained by Jones \cite{Jones}. Calculating higher powers of matrix ${\hat R}$ by repeated application of the Hecke condition
(\ref{Hecke1}), we find
\be {\hat R}^n = F_n^{(k)} {\hat R} + F_{n-1}^{(k)} I,\ee
in terms of Fibonacci divisors.

\section{Fibonacci Divisors}

\subsection{Addition and Division of Fibonacci Numbers}

The addition formula for Fibonacci numbers is given by following proposition.
\begin{prop} [Addition formula]
\begin{eqnarray}
F_{n+m}=F_{m}F_{n+1}+F_{n}F_{m-1}, \hskip0.5cm \mbox{where}\hskip0.5cm m,n \in \cal{Z} \label{additionformula}
\end{eqnarray}
\end{prop}
The proof is straightforward by using Binet formula for Fibonacci numbers.
If in (\ref{additionformula}) we put $m=n \equiv k$, then Fibonacci numbers with even index $n=2k$ can be factorized as
\bea
{F_{2k}=F_{k}\phantom{.} L_{k}},
\eea
where $L_{k} = F_{k-1}+F_{k+1}$ are the Lucas numbers.
 By repeating application of this formula
 \bea
F_{3k}&=&F_{k+2k}=F_{k}\phantom{.}F_{2k-1}+F_{k+1}\phantom{.}(F_{2k})=  \nonumber \\ & =& F_{k}\phantom{.}F_{2k-1}+F_{k+1}\phantom{.}(F_{k}\phantom{.} L_{k}) =F_{k}(F_{2k-1}+F_{k+1}\phantom{.} L_{k}) ,\nonumber
\eea
 we get factorization
 \bea
{F_{3k}=F_{k}\phantom{a}(F_{2k-1}+F_{k+1}\phantom{.}L_{k})}.
\eea
This can be continued
\bea
F_{4k}&=&F_{2k+2k}=F_{2k}\phantom{.}F_{2k-1}+F_{2k+1}\phantom{.}F_{2k}=F_{2k}(F_{2k-1}+F_{2k+1})= \\ &=&F_{2k}\phantom{a}L_{2k} =F_{k}\phantom{a}L_{k}\phantom{a}L_{2k}, \nonumber
\eea
with following factorization,
\bea
{F_{4k}=F_{k}\phantom{a}L_{k}\phantom{a}L_{2k}}.
\eea
The above results suggest the following divisibility property of $F_{n k}$.
\begin{prop}
 $F_{nk}$ is divisible by $F_{k}$.
\end{prop}
\begin{prf}
The proof can be done by induction on $n$. For given $n$, suppose ${F_{nk}=F_{k}\phantom{.}X(k,n)}$, where $X(k,n)\in \cal{Z}.$
Then, for $n+1$, by using $(\ref{additionformula})$ we have:
\bea
F_{(n+1)k}&=&F_{k+nk}=F_{k}\phantom{.} F_{nk-1}+\left[F_{nk}\right]\phantom{.} F_{k+1} \nonumber  \\
&=&F_{k} \phantom{.}F_{nk-1}+\phantom{.}\left[{F_{k}}\phantom{.}{X(k,n)}\right]\phantom{.}F_{k+1}\nonumber
=F_{k}( \phantom{.}F_{nk-1}+\phantom{.} X(k,n)\phantom{.}F_{k+1}).
\eea
 \end{prf}

\subsection{Fibonacci divisors conjugate to $F_k$}

Since $F_{nk}$ is divisible by $F_{k}$, the ratio $\frac{F_{nk}}{F_{k}}$ is an integer number.  We call this number as
Fibonacci divisor, conjugate to $F_k$.
\begin{definition}The Fibonacci divisor, conjugate to $F_k$ is an integer number
\bea
F_n^{(k)}=\frac{F_{nk}}{F_{k}}.
\eea
\end{definition}

\begin{prop}
The Binet type formula for Fibonacci divisor is,
\bea
F_n^{(k)}=\frac{(\varphi^k)^n-(\varphi'^k)^n}{\varphi^k-\varphi'^k}.
\eea
\end{prop}

\begin{prf}
It is derived simply by using the Binet formula for Fibonacci numbers,
\begin{eqnarray}
F_{nk}=\frac{(\varphi^k)^n-(\varphi'^k)^n}{\varphi-\varphi'}= \frac{(\varphi^k)^n-(\varphi'^k)^n}{\varphi^k-\varphi'^k} \frac{\varphi^k-\varphi'^k}{\varphi-\varphi'}= \frac{(\varphi^k)^n-(\varphi'^k)^n}{\varphi^k-\varphi'^k} \phantom{a} F_{k},\nonumber
\end{eqnarray}
so that
\begin{eqnarray}
\frac{F_{nk}}{F_k} = {F_n^{(k)}=\frac{(\varphi^k)^n-(\varphi'^k)^n}{\varphi^k-\varphi'^k}.}
\end{eqnarray}
\end{prf}

The Fibonacci divisor numbers $F_{n}^{(k)}$ give  factorization of Fibonacci numbers with factorized index $n k$:
\begin{eqnarray}
F_{nk}=F_{k}\phantom{a} F_{n}^{(k)}. \nonumber
\end{eqnarray}
The first few  sequences of Fibonacci divisors $F_n^{(k)}$ for $k=1,2,3,4,5$ and $n=1,2,3,4,5,...$ are given below
\begin{eqnarray}
 k &=& 1 ; F_n^{(1)} = F_n = 1,1,2,3,5,...\\
 k& = &2 ; F_n^{(2)} = F_{2n} = 1,3,8,21,55,...\\
 k&= &3; F_n^{(3)}= \frac{1}{2} F_{3n}= 1,4,17,72, 305,...\\
 k&=&4; F_n^{(4)} = \frac{1}{3} F_{4n}= 1,7,48,329,2255,...\\
 k&=&5; F_n^{(5)} = \frac{1}{5} F_{5n}=1,11,122,1353,15005,...
\end{eqnarray}

The next theorem gives the three terms  recursion formula  for Fibonacci divisors.
\begin{thm} (Recurrence relation for Fibonacci divisors)
\begin{eqnarray}
F_{n+1}^{(k)}=L_k F_{n}^{(k)}+(-1)^{k-1} F_{n-1}^{(k)}. \label{maintheorem}
\end{eqnarray}
\end{thm}
Here, $L_k$ are Lucas numbers. The formula is particular case of more general relation, given by following theorem.
\begin{thm}
\bea
F_{k(n+1)+\alpha}=L_k F_{kn+\alpha}+(-1)^{k-1}F_{k(n-1)+\alpha},\phantom{..} \mbox{where \phantom{.} $\alpha=0,1,\ldots,k-1.$}\label{mainalpha}
\eea
\end{thm}
The proof of both theorems is given in Appendix.

The total set of Fibonacci numbers $\{F_{n}\}$ is the sum of subsets $\{F_{kn+\alpha}\}$ for each k and $\alpha=0,1,\ldots,k-1$;\\
${k=2;} \phantom{..}\{F_{2 n},\phantom{.}F_{2 n+1}\}$ \\
${k=3;} \phantom{..}\{F_{3 n},\phantom{.}F_{3 n+1},\phantom{.}F_{3 n+2}\}$ \\
$\vdots$  \\
${k=k;} \phantom{..}\{F_{k n},\phantom{.}F_{k n+1},\ldots,\phantom{.}F_{k n+(k-1)}\}$ \\
Then equation $(\ref{mainalpha})$ state that for given k, the subsequences $F_{kn},\phantom{.}F_{kn+1},\phantom{.}F_{kn+2},\ldots,$ $F_{kn+(k-1)}$, satisfy the same recursion formula.
As an example
we consider following sequences with $k=3$;\\
\textbf{{$\alpha=0$}} \phantom{..} $\Rightarrow$ \phantom{..} $F_{3n}=0,2,8,34,\ldots$ \\
\textbf{{$\alpha=1$}} \phantom{..} $\Rightarrow$ \phantom{..} $F_{3n+1}=1,3,13,55,\ldots$ \\
\textbf{{$\alpha=2$}} \phantom{..} $\Rightarrow$ \phantom{..} $F_{3n+2}=1,5,21,89,\ldots$. \\
These sequences satisfy the same recursion formula,
\bea
F_{3(n+1)+\alpha}=L_3 F_{3n+\alpha}+(-1)^{3-1}F_{3(n-1)+\alpha},  \nonumber
\eea
where $\alpha=0,1,2$,
but with different initial values, so that their union set covers the whole Fibonacci sequence.
The corresponding
Fibonacci divisors $F_n^{(3)}$ conjugate to $F_3$
are determined by recursion with Lucas number $L_3 = 4$,
$$    F_{n+1}^{(3)}= 4 F_n^{(3)}+ F_{n-1}^{(3)} ,                    $$
and the initial values $F_0^{(3)}=0$ and $F_1^{(3)}=1$. This gives the  sequence of numbers $0, 1,  4, 17, 72, 305,...$

\begin{prop}
The Fibonacci divisors\phantom{.}$F_{n}^{(k)}$
can be extended to negative integers according to formulas
\bea
F_{-n}^{(k)}&=&(-1)^{k n+1}\phantom{.} F_{n}^{(k)}, \\
F_{n}^{(-k)}&=&(-1)^{(n+1)k}\phantom{.} F_{n}^{(k)}, \label{higherorderfibnumberswithminusk} \\
F_{-n}^{(-k)}&=&(-1)^{k+1} \phantom{.} F_{n}^{(k)}.
\eea
\end{prop}
These formulas can be proved by direct application of the Binet representation.
The formulas determine $F_{n}^{(k)}$ for each $k \in \cal{Z}$, and each $n \in \cal{Z}$.

The following proposition relates powers of the Golden ratio with Fibonacci divisors.
\begin{prop} For $k\in Z$ and $n \in Z$,
\bea
(\varphi^k)^n&=&\varphi^k \phantom{.} F_{n}^{(k)}+(-1)^{k+1} \phantom{.}F_{n-1}^{(k)}, \label{goldenpowerkn} \\
(\varphi'^k)^n&=&\varphi'^k \phantom{.} F_{n}^{(k)}+(-1)^{k+1} \phantom{.} F_{n-1}^{(k)}. \label{silverpowerkn}
\eea
\end{prop}
The proof is straightforward.

\begin{cor}
Numbers $\varphi^k$ and ${\varphi'}^k$ are roots of quadratic characteristic equation
\be
\lambda_k^2 = L_k \lambda_k + (-1)^{k+1}, \nonumber
\ee
so that
\be
\left(\varphi^k\right)^2 = L_k \varphi^k + (-1)^{k+1},\,\,\,\, \left({\varphi'}^k\right)^2 = L_k {\varphi'}^k + (-1)^{k+1}.\nonumber
\ee
\end{cor}
The last relations follow from the proposition for $n=2$ and $F_2^{(k)} = \varphi^k + {\varphi'}^k = L_k$.

\section{Hierarchy of Golden Derivatives}
In this section we introduce the Golden derivative operators $_{(k)} D^{x}_{F}$ corresponding to Fibonacci divisors, conjugate to $F_k$.

\begin{definition} For arbitrary function $f(x)$ and $k \in Z$,
\bea
_{(k)} D^{x}_{F}[f(x)]=\frac{f(\varphi^{k} x)-f(\varphi'^{k} x)}{\left(\varphi^{k} -\varphi'^{k}\right)x}. \label{definitionofhigherorderderivative}
\eea
\end{definition}

The linear operator $_{(k)} D^{x}_{F}$ we call the $k^{th}$ Golden derivative operator.
The complex analytic version of this derivative, acting on an analytic function $f(z)$, can be represented by the Fibonacci
divisor number operator  $F^{(k)}_N$ in the Fock-Bargman representation, where  $N= z \frac{d}{dz}$,
\be
F^{(k)}_{z \frac{d}{dz}} f(z)
= \frac{\varphi^{k z \frac{d}{dz}} - {\varphi'}^{k z \frac{d}{dz}}}{\varphi^k - {\varphi'}^k} f(z)
= z \,_{(k)} D^{z}_{F}[f(z)] .
\ee
For even $k$ this formula admits the limit $k \rightarrow 0$, giving the usual derivative
\be
\lim_{k \rightarrow 0} D^{z}_{F}[f(z)]  = f'(z).
\ee
For $k=1$, the derivative reduces to the Golden derivative \cite{golden},
\bea
_{(1)} D^{x}_{F}[f(x)]=D^{x}_{F}[f(x)] = \frac{f(\varphi x)-f(\varphi' x)}{\left(\varphi -\varphi'\right)x},
\eea
which by acting on monomial $x^n$ produces Fibonacci numbers,
\begin{eqnarray}
_{(1)} D^{x}_{F}(x^n)=\frac{(\varphi x)^n-(\varphi' x)^n}{(\varphi-\varphi')x}=\frac{{\varphi }^n-{\varphi'}^n}{\varphi-\varphi'}x^{n-1}=F_n\phantom{a}x^{n-1}. \nonumber
\end{eqnarray}
In a similar way, by applying the $k^{th}$ Golden derivative  to  $x^n$, we get the Fibonacci divisor numbers $F_n^{(k)}$, conjugate to $F_k$,
\bea
_{(k)} D^{x}_{F}(x^n)=\frac{(\varphi^k x)^n-(\varphi'^k x)^n}{(\varphi^k -\varphi'^k )x}=\frac{(\varphi^k)^n-(\varphi'^k)^n}{\varphi^k -\varphi'^k} x^{n-1}=F_n^{(k)} x^{n-1}, \nonumber
\eea
or
\bea
_{(k)} D^{x}_{F}[x^n]=F_n^{(k)} x^{n-1}.  \label{kthderivativeapplicationtoxpowern}
\eea
This formula for negative $k$  produces numbers, in accordance with $(\ref{higherorderfibnumberswithminusk})$,
\bea
F_{n}^{(-k)}&=&(-1)^{(n+1)k}\phantom{.} F_{n}^{(k)}.      \nonumber
\eea

 The Leibnitz and quotient rules for $k^{th}$ Golden derivative are subject of next propositions.

\begin{prop} (The Leibnitz Rule)
\begin{eqnarray}
_{(k)} D^{x}_{F}(f(x)g(x))= _{(k)} D^{x}_{F}(f(x))\phantom{.}g(\varphi^k x)+f\left(\varphi'^k x \right)\phantom{.}_{(k)} D^{x}_{F}(g(x)). \label{kthderivativeleibnitz}
\end{eqnarray}
\end{prop}

\begin{prop} (The Quotient Rule)
\begin{eqnarray}
_{(k)} D^{x}_{F}\left(\frac{f(x)}{g(x)}\right)=\frac{_{(k)} D^{x}_{F}(f(x)) \phantom{.} g(\varphi^k x)-f(\varphi^k  x) \phantom{.} _{(k)}  D^{x}_{F}(g(x))}{g(\varphi^k  x)\phantom{.}g\left(\varphi'^k x\right)}.
\end{eqnarray}
\end{prop}
By applying Leibnitz rule to (\ref{kthderivativeapplicationtoxpowern}),
\bea
_{(k)} D^{x}_{F}(x^n)=_{(k)} D^{x}_{F}(x^m x^{n-m}) = F_n^{(k)} x^{n-1},\label{comparingresult1}
\eea
we get the following corollary.

\begin{cor}
\bea
{F_n^{(k)}=F_m^{(k)} \phantom{.}\varphi^{k(n-m)}+F_{n-m}^{(k)} \phantom{.}{\varphi'}^{k m}}.\label{corol}
\eea
\end{cor}
This corollary allows us to formulate following proposition.
\begin{prop}
\bea
F_n^{(k)}=F_{n-m}^{(k)}\phantom{.}F_{m+1}^{(k)}+(-1)^{k+1} \phantom{.}F_m^{(k)}\phantom{.}F_{n-m-1}^{(k)}.\label{adition}
\eea
In particular case $k=1$ it gives addition formula for Fibonacci numbers,
\bea
F_n=F_{n-m}\phantom{.}F_{m+1}+F_m\phantom{.}F_{n-m-1}. \nonumber
\eea
\end{prop}
\begin{prf}
By substituting $(\ref{goldenpowerkn})$ and $(\ref{silverpowerkn})$, to  (\ref{corol}) we get
\bea
F_n^{(k)}&=&F_m^{(k)} \phantom{.}\left(\varphi^k F_{n-m}^{(k)} +(-1)^{k+1}F_{n-m-1}^{(k)} \right)+F_{n-m}^{(k)} \phantom{.}\left( \varphi'^k F_{m}^{(k)} +(-1)^{k+1}F_{m-1}^{(k)}   \right) \nonumber \\
&=&F_m^{(k)}\phantom{.} F_{n-m}^{(k)}\left(\varphi^k+\varphi'^k\right)+(-1)^{k+1} F_m^{(k)} \phantom{.} F_{n-m-1}^{(k)}+(-1)^{k+1} F_{n-m}^{(k)} \phantom{.} F_{m-1}^{(k)} \nonumber \\
&=&F_m^{(k)}\phantom{.} F_{n-m}^{(k)}\phantom{.} L_{k}+(-1)^{k+1} F_m^{(k)} \phantom{.} F_{n-m-1}^{(k)}+(-1)^{k+1} F_{n-m}^{(k)} \phantom{.} F_{m-1}^{(k)} \nonumber \\
&=&F_{n-m}^{(k)} \left(L_{k} \phantom{.} F_m^{(k)} +(-1)^{k+1}\phantom{.} F_{m-1}^{(k)}\right)+(-1)^{k+1} F_m^{(k)} \phantom{.} F_{n-m-1}^{(k)} \nonumber \\
&{(\ref{maintheorem})}{=}&F_{n-m}^{(k)}\phantom{.}F_{m+1}^{(k)}+(-1)^{k+1} \phantom{.}F_m^{(k)}\phantom{.}F_{n-m-1}^{(k)}.  \nonumber
\eea
\end{prf}

By shifting index $n$ in (\ref{adition}) we get addition formula for $F_{n}^{(k)}$in convenient form.
\begin{prop} An addition formula for Fibonacci divisor numbers is
\bea
F_{n+m}^{(k)}=F_{m}^{(k)}\phantom{.}F_{n+1}^{(k)}+(-1)^{k+1} \phantom{.}F_n^{(k)}\phantom{.}F_{m-1}^{(k)}. \nonumber
\eea
For $k=1$, it reduces to addition formula  for Fibonacci numbers (\ref{additionformula}).
\end{prop}

\subsection{Hierarchy of Golden Periodic Functions}

The Golden periodic function for Golden derivative plays the same role as a constant for the usual derivative.
Similarly, the set of $k^{th}$ Golden derivatives (\ref{definitionofhigherorderderivative}) determines the hierarchy of Golden periodic functions for
every natural $k$.
\begin{prop} \label{propofgoldenperiodicfunction} Every Golden periodic function $D^{x}_{F}(f(x))=0$ $\Longleftrightarrow$ $(f(\varphi x)=f(\varphi' x))$ \phantom{.} is also periodic for arbitrary $k-th $ order Golden derivatives,
\bea
 D^{x}_{F}(f(x))=0 \,\,\,\,&\Rightarrow& \,\,\,\, _{(k)} D^{x}_{F}(f(x))=0,\nonumber\\
 f(\varphi x)=f(\varphi' x)\,\,\,\,&\Rightarrow& \,\,\,\, f(\varphi^k x)=f(\varphi'^k x),
\eea
 for $k=2,3,\ldots$
\end{prop}
The proof is evident by induction.
As an example, $f(x)=\sin\left(\frac{\pi}{\ln \varphi}\ln |x|\right)$ as the Golden periodic function,
\bea
D^{x}_{F} \sin\left(\frac{\pi}{\ln \varphi}\ln |x|\right)   = 0,    \nonumber
\eea
is  also the $k-th$ Golden periodic,
  \bea
_{(k)} D^{x}_{F} \sin\left(\frac{\pi}{\ln \varphi}\ln |x|\right)   =0 . \nonumber
\eea
But the opposite to Proposition $ \ref{propofgoldenperiodicfunction}$ is not in general true. If function $f(x)$ is $k$- periodic, $k=2,3...$,
\bea
_{(k)} D^{x}_{F}(f(x))=0, \nonumber
\eea
then it is not necessarily the Golden periodic one (with $k=1$).
For example,  $f(x)= \sin\left(\frac{\pi}{\ln \varphi^2} \ln |x|\right)$ is the Golden periodic function with $k=2, $ i.e\phantom{..} $_{(2)} D^{x}_{F}(f(x))=0$. However,
the Golden derivative of this function doesn't vanish,
\bea
D^{x}_{F}(f(x))=2 \phantom{.} \frac{\cos\left(\frac{\pi}{\ln (\varphi^2)} \ln |x| \right)}{(\varphi-\varphi')x}\neq 0  \nonumber
\eea
and it is not the Golden periodic function.

\subsection{Generating Function for Fibonacci divisors}

\begin{definition}
Function,
\bea
_{(k)} F(x)=\sum_{n=0}^\infty F_n^{(k)} x^n = F_0^{(k)}+F_1^{(k)} x+F_2^{(k)} x^2+F_3^{(k)} x^3+\ldots \nonumber
\eea
is called the generating function for Fibonacci divisors $F_n^{(k)}$, where
\bea
F_n^{(k)}=\frac{1}{n!} \frac{d^n}{d x^n} \phantom{.} {_{(k)} F(x) }\phantom{.}\bigg|_{x=0}.
\eea

\end{definition}

\begin{prop}
The generating function $_{(k)} F(x)$ is convergent in the disk $|x|<\frac{1}{\varphi^k}$ and has following explicit representation
\bea
_{(k)} F(x)=\sum_{n=0}^\infty F_n^{(k)} x^n =\frac{1}{1-L_{k} x+(-1)^k x^2}.
\eea
\end{prop}

\begin{prf}
The ratio test
\bea
\rho=\lim_{n{\to \infty}} \left| \frac{F_{n+1}^{(k)} x^{n+1}}{F_n^{(k)} x^n} \right|=\lim_{n{\to \infty}} \left| \frac{F_{n+1}^{(k)}}{F_{n}^{(k)}} \right| \phantom{.}|x|=\varphi^k \phantom{.} |x| < 1 \nonumber
\eea
implies,
\bea
|x| < \left(\frac{1}{\varphi}\right)^k < 1,\nonumber
\eea
for any positive $k$. By using $k-th$ Golden derivative
\bea
_{(k)} F(x)&=&\sum_{n=0}^\infty F_n^{(k)} x^n =F_0^{(k)}+ \sum_{n=1}^\infty x F_n^{(k)} x^{n-1} =0+ \sum_{n=1}^\infty x \phantom{.}F_n^{(k)} \phantom{.}x^{n-1}{=} \nonumber \\ &=&\sum_{n=1}^\infty x \phantom{.} {_{(k)} D^{x}_{F}(x^n)}
=x \phantom{.}_{(k)} D^{x}_{F} \sum_{n=1}^\infty x^n =\nonumber \\
&=&x \phantom{.}_{(k)} D^{x}_{F}\phantom{.}x\left(1+x+x^2+\ldots\right)\nonumber \\
&{=}& x \phantom{.} _{(k)} D^{x}_{F}\left(\frac{x}{1-x} \right) =x\left[_{(k)} D^{x}_{F}\left(\frac{1}{1-x} \right) \right] = x \phantom{.} \frac{\left(\frac{1}{1-\varphi^k x}-\frac{1}{1-\varphi'^k x}\right)  }{(\varphi^k-\varphi'^k)x}= \nonumber \\
&=&\frac{x}{(1-\varphi^k x)(1-\varphi'^k x)} = \frac{x}{1-L_k \phantom{.}x+\left(-1\right)^k x^2}. \nonumber
\eea
\end{prf}
\begin{cor}
$_{(k)} F(x)$ is rational function with simple zero at $x=0$ and two simple poles at
$$x=\frac{1}{\varphi^k}, \phantom{...}x=\frac{1}{\varphi'^k}.$$
\end{cor}
For
${k=1}$ it reduces to generating function for Fibonacci numbers,
\bea
_{(1)} F(x)= \sum_{n=0}^\infty F_n x^n=\frac{x}{1-x-x^2}=\frac{x}{(1-\varphi x)(1-\varphi' x)}.
\eea
For  ${k=2}$ and ${k=3}$ it gives generating function for "mod 2" and "mod 3" Fibonacci numbers correspondingly,
\bea
_{(2)} F(x)=\sum_{n=0}^\infty F_{2n} x^n=\frac{x}{1-3x+x^2}=\frac{x}{(1-\varphi^2 x)(1-\varphi'^2 x)},
\eea

\bea
_{(3)} F(x)=\frac{1}{2} \sum_{n=0}^\infty F_{3n} x^n=\frac{x}{1-4x-x^2}=\frac{x}{(1-\varphi^3 x)(1-\varphi'^3 x)}.
\eea
For arbitrary $k$ it represents generating function for "mod $k$"-Fibonacci numbers,
\bea
_{(k)} F(x)=\frac{1}{F_{k}} \sum_{n=0}^\infty F_{kn} x^n=\frac{x}{1-L_{k} x+(-1)^k x^2}=\frac{x}{(1-\varphi^k x)(1-\varphi'^k x)}.
\eea

\section{Entire generating function for Fibonacci divisors}

Applying $_{(k)} D^{x}_{F}$ to $e^{x}$ in power series form
\begin{eqnarray}
_{(k)} D^{x}_{F}(e^{x})&=&_{(k)} D^{x}_{F} \left(\sum_{n=0}^\infty \frac{x^n}{n!}\right)=_{(k)} D^{x}_{F}\left(\frac{1}{0!}+\frac{x}{1!}+\frac{x^2}{2!}+\frac{x^3}{3!}+... \right) \nonumber \\
&=& \sum_{n=1}^\infty \frac{_{(k)} D^{x}_{F}(x^n)}{n!}=\sum_{n=1}^\infty F^{(k)}_{n} \frac{x^{n-1}}{n!}=\sum_{n=0}^\infty F^{(k)}_{n+1} \frac{x^{n}}{(n+1)!} \nonumber
\end{eqnarray}
we get the series
\begin{eqnarray}
_{(k)} D^{x}_{F}(e^{x})=\sum_{n=0}^\infty F^{(k)}_{n+1} \frac{x^{n}}{(n+1)!}, \label{higher.der.of.exp.f.1}
\end{eqnarray}
convergent for arbitrary $x$.  From another side, by definition $(\ref{definitionofhigherorderderivative})$,
\begin{eqnarray}
_{(k)} D^{x}_{F}(e^{x})=\frac{e^{\varphi^k x}-e^{\varphi'^k x}}{(\varphi^k-\varphi'^k)x}
=e^{\frac{\varphi^k +\varphi'^k }{2}x}\frac{e^{\left(\frac{\varphi^k-\varphi'^k}{2}\right) x}-e^{-\left(\frac{\varphi^k-\varphi'^k}{2}\right) x}}{\left(\varphi^k-\varphi'^k\right)x}, \nonumber
\end{eqnarray}
and relations $\varphi^k+\varphi'^k=L_k$, $\varphi^k-\varphi'^k=F_k(\varphi-\varphi')$, we have
\begin{eqnarray}
_{(k)} D^{x}_{F}(e^{x})=2 e^{\frac{L_k}{2}x} \phantom{.}  \frac{\sinh \left(\frac{F_k}{2}(\varphi-\varphi')x\right)}{F_k (\varphi-\varphi') x}. \nonumber
\end{eqnarray}
Replacing $\varphi-\varphi'=\sqrt{5}$, we get explicit formula for derivative of exponential function
\begin{eqnarray}
_{(k)} D^{x}_{F}(e^{x})= e^{\frac{L_k}{2}x} \phantom{.}  \frac{\sinh \left(F_k\frac{\sqrt{5}}{2}x\right)}{\left(F_k \frac{\sqrt{5}}{2} x\right)} . \label{higher.der.of.exp.f.2}
\end{eqnarray}
By comparing (\ref{higher.der.of.exp.f.1}) and (\ref{higher.der.of.exp.f.2}), finally we obtain identity:
\begin{eqnarray}
{\sum_{n=1}^\infty   \frac{F^{(k)}_{n}}{n!} x^n = e^{\frac{L_k}{2}x} \phantom{.}  \frac{\sinh \left(F_k\frac{\sqrt{5}}{2}x\right)}{F_k \frac{\sqrt{5}}{2} }  }. \label{exphigherderivativeequality}
\end{eqnarray}
In particular case $k=1$, it reduces to the one for Fibonacci numbers \cite{golden},
\begin{eqnarray}
{\sum_{n=1}^\infty   \frac{F_{n}}{n!} x^n = e^{\frac{1}{2}x} \phantom{.}  \frac{\sinh \left(\frac{\sqrt{5}}{2}x\right)}{\frac{\sqrt{5}}{2} }  }.
\end{eqnarray}
Relation (\ref{exphigherderivativeequality}) represents the entire generating function for Fibonacci divisors,
and allows us to obtain several interesting identities.

\subsection{Some identities for $F^{(k)}_{n}$}
From (\ref{exphigherderivativeequality}) for $x=1$ follow summation formulas
\bea
\sum_{n=0}^\infty   \frac{F^{(k)}_{n}}{n!} = e^{\frac{L_k}{2}} \phantom{.}  \frac{\sinh \left(F_k\frac{\sqrt{5}}{2}\right)}{\left(F_k \frac{\sqrt{5}}{2}\right)},
\eea
and
\bea
\sum_{n=0}^\infty   \frac{F_{nk}}{n!} = e^{\frac{L_k}{2}} \phantom{.}  \frac{\sinh \left(F_k\frac{\sqrt{5}}{2}\right)}{\left(\frac{\sqrt{5}}{2}\right)} .
\eea
  By replacing $x \rightarrow i x$ in ($\ref{exphigherderivativeequality}$) we get,
\bea
\sum_{n=1}^\infty   \frac{F^{(k)}_{n}}{n!} (i)^{n} x^n = e^{i\frac{L_k}{2}x} \phantom{.}  \frac{\sinh \left(F_k\frac{\sqrt{5}}{2}i x\right)}{ F_k \frac{\sqrt{5}}{2}}.
\eea
Substituting $\sinh(i x)=i \sin(x)$  and splitting  sum in the l.h.s. to even and odd parts gives,
\bea
\sum_{l=0}^\infty   \frac{F^{(k)}_{2l+1}}{(2l+1)!} (-1)^{l} x^{2l+1}+i \sum_{l=0}^\infty   \frac{F^{(k)}_{2l+2}}{(2l+2)!} (-1)^{l} x^{2l+2} = e^{i\frac{ L_k}{2}x} \phantom{..}\frac{\sin(F_k \frac{\sqrt5}{2} x)}{F_k\frac{\sqrt{5}}{2} }
\eea
By using Euler formula  $e^{i\frac{L_k}{2}x}=\cos(\frac{L_k}{2}x) +i \sin(\frac{L_k}{2}x)$,
 and splitting equality to real and imaginary parts, we get generating functions for even and odd Fibonacci divisors:
\bea
{\sum_{l=0}^\infty   \frac{F^{(k)}_{2l+1}}{(2l+1)!} (-1)^{l} x^{2l+1}= \cos\left(\frac{L_k}{2}x\right) \phantom{.} \frac{\sin(F_k \frac{\sqrt5}{2} x)}{F_k\frac{\sqrt{5}}{2} }}, \label{realpartidentityforhigherderivative}
\eea
and
\bea
{\sum_{l=0}^\infty   \frac{F^{(k)}_{2l+2}}{(2l+2)!} (-1)^{l} x^{2l+2} =\sin\left(\frac{L_k}{2}x\right) \frac{\sin(F_k \frac{\sqrt5}{2} x)}{F_k\frac{\sqrt{5}}{2} }}. \label{imaginarypartidentityforhigherderivative}
\eea
From these entire functions follow several identities. From $(\ref{realpartidentityforhigherderivative})$ we have:\\
$1)\phantom{.}\mbox{For} \phantom{.} x=\pi$,\\
\bea
\sum_{l=0}^\infty   \frac{F^{(k)}_{2l+1}}{(2l+1)!} (-1)^{l} {\pi}^{2l}= \cos\left(\frac{\pi}{2}L_k\right) \phantom{.} \frac{\sin(F_k \frac{\sqrt5}{2} \pi)}{F_k\frac{\sqrt{5}}{2} \pi}.
\eea
The right hand side of this identity vanishes for odd values of Lucas numbers $L_k$: $L_1 =1, L_2 =3, L_4 = 7, L_5 = 11, etc$. \\
$2)\phantom{..} x=\frac{2 \pi}{\sqrt{5}}$,\\
Since $\sin(F_k \pi)=0$, then for arbitrary $k$;
\bea
\sum_{l=0}^\infty   \frac{F^{(k)}_{2l+1}}{(2l+1)!} \phantom{.}(-1)^{l} \phantom{.}\frac{(2 \pi)^{2l}}{5^l}=0.
\eea
$3)\phantom{..} x=\frac{\pi}{\sqrt{5}}$,
\bea
\sum_{l=0}^\infty   \frac{F^{(k)}_{2l+1}}{(2l+1)!} (-1)^{l} \frac{\pi^{2l}}{5^l}= \frac{2}{F_k \pi} \cos\left(L_k \frac{\pi}{2\sqrt{5}}\right) \phantom{.} \sin\left(F_k \frac{\pi}{2}\right).
\eea
For even values of $F_k$; $F_3 = 2, F_6 = 8, F_9 = 34, etc.$ the right hand side vanishes. \\
$4)\phantom{..} x=2 \pi$,
\bea
\sum_{l=0}^\infty   \frac{F^{(k)}_{2l+1}}{(2l+1)!} (-1)^{l} {(2 \pi)}^{2l}= \cos(L_k \pi) \phantom{.} \frac{\sin(F_k \sqrt{5} \pi)}{F_k \sqrt{5} \pi}.
\eea
$5)\phantom{..} x=1$,
\bea
\sum_{l=0}^\infty   \frac{F^{(k)}_{2l+1}}{(2l+1)!} (-1)^{l}= \cos\left(\frac{L_k}{2}\right) \phantom{.}  \frac{\sin(F_k \frac{\sqrt5}{2})}{F_k\frac{\sqrt{5}}{2}}.
\eea
$6)\phantom{..} x=\frac{\pi}{L_k}$,
\bea
\sum_{l=0}^\infty   \frac{F^{(k)}_{2l+1}}{(2l+1)!} (-1)^{l} \left(\frac{\pi}{L_k}\right)^{2l}=0.
\eea
$7)\phantom{..} x=\frac{2 \pi}{\sqrt{5} F_k}$,
\bea
\sum_{l=0}^\infty   \frac{F^{(k)}_{2l+1}}{(2l+1)!} (-1)^{l} \frac{{(2 \pi)}^{2l}}{5^l (F_k)^{2l}}=0.
\eea
In a similar way from $(\ref{imaginarypartidentityforhigherderivative})$ follow identities:\\
$1)\phantom{..} x=\pi$,
\bea
\sum_{l=0}^\infty   \frac{F^{(k)}_{2l+2}}{(2l+2)!} (-1)^{l} {(\pi)}^{2l+1}= \sin\left(\frac{\pi}{2}L_k \right) \phantom{.} \frac{\sin(F_k \frac{\sqrt5}{2} \pi)}{F_k\frac{\sqrt{5}}{2} \pi}.
\eea
For even Lucas numbers: $L_3 = 4, L_6 = 18, L_9 = 76, etc.$ the right hand side is zero.\\
$2)\phantom{..} x=\frac{2 \pi}{\sqrt{5}}$. Since $\sin(F_k \pi)=0$, then
\bea
\sum_{l=0}^\infty   \frac{F^{(k)}_{2l+2}}{(2l+2)!} \phantom{.}(-1)^{l} \phantom{.}\left(\frac{2 \pi}{\sqrt{5}}\right)^{2l+1}=0.
\eea
$3)\phantom{..} x=\frac{\pi}{\sqrt{5}}$,
\bea
\sum_{l=0}^\infty   \frac{F^{(k)}_{2l+2}}{(2l+2)!} (-1)^{l} \left(\frac{\pi}{\sqrt{5}}\right)^{2l+1}= \frac{2}{F_k \pi} \sin \left(L_k \frac{\pi}{2\sqrt{5}} \right) \phantom{.} \sin \left(F_k \frac{\pi}{2} \right).
\eea
For even Fibonacci numbers $F_k$: $F_3 = 2, F_6 = 8, F_9 = 34, etc.$ the right hand side vanishes. \\
$4)\phantom{..} x=2 \pi$,
\bea
\sum_{l=0}^\infty   \frac{F^{(k)}_{2l+2}}{(2l+2)!} (-1)^{l} {(2 \pi)}^{2l+1}= 0.
\eea
$5)\phantom{..} x=1$,
\bea
\sum_{l=0}^\infty   \frac{F^{(k)}_{2l+2}}{(2l+2)!} (-1)^{l}= \sin \left(\frac{L_k}{2} \right) \phantom{.}  \frac{\sin(F_k \frac{\sqrt5}{2})}{F_k\frac{\sqrt{5}}{2}}.
\eea
$6)\phantom{..} x=\frac{\pi}{L_k}$,
\bea
\sum_{l=0}^\infty   \frac{F^{(k)}_{2l+2}}{(2l+2)!} (-1)^{l} \left(\frac{\pi}{L_k}\right)^{2l+1}=\frac{\sin\left( \frac{F_k}{L_k}\frac{\sqrt{5}}{2} \pi\right)}{ \frac{F_k}{L_k}\frac{\sqrt{5}}{2} \pi}.
\eea
$7)\phantom{..} x=\frac{2 \pi}{L_k}$. Since $\sin(\frac{L_k}{2}\frac{2\pi}{L_k})=\sin(\pi)=0$, then
\bea
\sum_{l=0}^\infty   \frac{F^{(k)}_{2l+2}}{(2l+2)!} (-1)^{l} \left(\frac{2 \pi}{L_k}\right)^{2l+1}=0.
\eea
For $x=\frac{\pi}{\sqrt{5} F_k}$, two more interesting identities follow from $(\ref{realpartidentityforhigherderivative})$ and $(\ref{imaginarypartidentityforhigherderivative})$ correspondingly,
\bea
\sum_{l=0}^\infty   \frac{F^{(k)}_{2l+1}}{(2l+1)!} (-1)^{l} \frac{{\pi}^{2l}}{5^l F^{2l}_k}=\frac{2}{\pi} \cos\left(\frac{L_k}{F_k} \frac{\pi}{2 \sqrt{5}}\right),
\eea
and
\bea
\sum_{l=0}^\infty   \frac{F^{(k)}_{2l+2}}{(2l+2)!} (-1)^{l} \left(\frac{\pi}{\sqrt{5} F_k}\right)^{2l+1}=\frac{2}{\pi} \sin\left(\frac{L_k}{F_k} \frac{\pi}{2 \sqrt{5}}\right).
\eea

\section{Fibonacci divisors and  Fibonomials}
\begin{definition}
The product,
\bea
k \cdot 2k \cdot 3k \ldots nk \equiv n!_{mod \phantom{.} k}
\eea
is called mod k factorial. It is equal,
\bea
\prod_{s=1}^{n} s k=n! k^n
\eea
and for particular case it reduces to usual factorial
\bea
k=1 &\Rightarrow& n!_{mod \phantom{.} 1}=n!  \nonumber
\eea
\end{definition}

\begin{definition} The product of Fibonacci numbers,
\bea
F_{k} F_{2k} \ldots F_{nk}= \prod_{s=1}^{n} F_{s k} \equiv {F_{n}!}_{\phantom{.}mod \phantom{.} k}
\eea
is called mod k Fibonacci factorial. For $k=1$, it gives the Fibonacci factorial,
\bea
{F_{n}!}_{\phantom{.}mod \phantom{.} 1}=F_{1} F_{2} \ldots F_{n}= F_{n}!
\eea
For $k=2$ and even $n$, it gives the double Fibonacci factorial,
\bea
{F_{n}!}_{\phantom{.}mod \phantom{.} 2}=F_{2} F_{4} \ldots F_{2n}.
\eea
\end{definition}

\begin{definition}
The product of Fibonacci divisors,
\bea
F_1^{(k)} F_2^{(k)} \ldots F_n^{(k)}=\prod_{i=1}^{n} F_i^{(k)} \equiv F_n^{(k)}!
\eea
is called the Fibonacci divisors factorial. It can be considered as the $k-th$ Fibonorial or generalized Fibonorial. In particular case $k=1$, it reduces to Fibonacci factorial; $F_n^{(1)}!=F_{n}!$. For  $F_n^{(k)}!$ we have next formula,
\bea
F_n^{(k)}!=\frac{F_{k} F_{2k} F_{3k} \ldots F_{nk}}{F_{k} F_{k} F_{k} \ldots F_{k}}=\frac{F_{k} F_{2k} F_{3k} \ldots F_{nk}}{\left(F_{k}\right)^n},
\eea
or in terms of mod k Fibonacci factorial,
\bea
F_n^{(k)}!=\frac{{F_{n}!}_{\phantom{.}mod \phantom{.} k}}{\left(F_{k}\right)^n}. \label{intermsofmodkFibonaccifactorial}
\eea
\end{definition}

\begin{definition}
The  Fibonomial coefficients for Fibonacci divisors or shortly $k-th$ Fibonomials are defined as
\bea
_{(k)}  {n \brack m}_{F}=\frac{F_1^{(k)} F_2^{(k)} \ldots F_{n-m+1}^{(k)}}{F_1^{(k)} F_2^{(k)} \ldots F_m^{(k)}}=\frac{F_n^{(k)}!}{F_m^{(k)}! F_{n-m}^{(k)}!},
\eea
and for $k=1$ they reduce to Fibonomials,
\bea
{n \brack m}_{F}=\frac{F_{n}!}{F_{n-m}! F_{m}!}. \nonumber
\eea
For arbitrary $k$ they can be represented by mod k Fibonacci factorials $(\ref{intermsofmodkFibonaccifactorial})$;
\bea
_{(k)} {n \brack m}_{F}=\frac{{F_{n}!}_{\phantom{.}mod \phantom{.} k}}{{F_{m}!}_{\phantom{.}mod \phantom{.} k} \phantom{.} {F_{n-m}!}_{\phantom{.}mod \phantom{.} k}}.
\eea
\end{definition}
In a similar way as for Fibonomials, it is possible to derive the recursion formula for $k-th$ Fibonomials
\bea
{_{(k)}} {n \brack m}_{F}= \varphi'^{km} {_{(k)}} {n-1 \brack m}_{F} +  \varphi^{k(n-m)}  {_{(k)}} {n-1 \brack m-1}_{F},  \label{recursionbinomial}
\eea
and give interpretation of it in terms of Pascal type triangle.

\section{Hierarchy of Golden binomials}
By $k-th$ Fibonomials we can  introduce now the hierarchy of Golden binomials.
\begin{definition}
The $k-th$ Golden binomial is defined by polynomial,
$$
_{(k)} \left(x-a\right)^{n}_{F} =\left( x-\varphi^{k(n-1)} a\right)\left( x-\varphi^{k(n-2)} \varphi'^k a\right)\ldots \left( x-\varphi^{k} \varphi'^{k(n-2)} a\right)\left( x-\varphi'^{k(n-1)} a\right)$$
if  $n \geq 1$, and it is equal one if $n=0$.
\end{definition}
In particular case $k=1$, it reduces to the Golden binomial \cite{golden}. The polynomial satisfies following factorization formula.
\begin{prop} (Factorization Property)
\bea
_{(k)} \left(x-a\right)^{n+m}_{F}&=& _{(k)} \left(x-\varphi^{km} a\right)^{n}_{F} \phantom{.}  _{(k)} \left(x-\varphi'^{kn} a\right)^{m}_{F} \\
&=& _{(k)} \left(x-\varphi'^{km} a\right)^{n}_{F} \phantom{.}  _{(k)} \left(x-\varphi^{kn} a\right)^{m}_{F}
\eea
\end{prop}
 The proof is straightforward.

\begin{thm}
The $k-th$ Golden binomial expansion is,
\bea
_{(k)} \left(x+y\right)^{n}_{F}=\sum_{m=0}^n    {_{(k)}} {n \brack m}_{F} (-1)^{k\frac{m(m-1)}{2}}x^{n-m} y^{m} \label{Highergoldenbinomialexpansion}
\eea
In particular case $k=1$, it reduces to the Golden binomial formula \cite{golden}.
\end{thm}
The proof can be done by induction.

\begin{cor}
From this theorem the identity follows,
\bea
_{(k)} \left(1+1\right)^{n}_{F}=\sum_{m=0}^n  {_{(k)}} {n \brack m}_{F} (-1)^{k\frac{m(m-1)}{2}}.
\eea
\end{cor}

\begin{lem}
The $k-th$ Golden derivatives are acting on $k-th$ Golden binomials as,
\bea
{_{(k)} D^{x}_{F}}\phantom{..} _{(k)} \left(x+y\right)^{n}_{F}&=& F_n^{(k)}\phantom{.}  _{(k)} \left(x+y\right)^{n-1}_{F},  \\
{_{(k)} D^{y}_{F}}\phantom{..} _{(k)} \left(x+y\right)^{n}_{F}&=& F_n^{(k)}\phantom{.}  _{(k)} \left(x+(-1)^k y\right)^{n-1}_{F},  \\
{_{(k)} D^{y}_{F}}\phantom{..} _{(k)} \left(x-y\right)^{n}_{F}&=&-F_n^{(k)}\phantom{.}  _{(k)} \left(x-(-1)^k y\right)^{n-1}_{F}.
\eea
\end{lem}
The proof is long but straightforward.

\subsection{Fibonacci divisors and Golden Taylor expansion}

Here we
introduce monomials,
\bea
P^{(k)}_n \equiv \frac{x^n}{F_n^{(k)}!},
\eea
so that due to $(\ref{kthderivativeapplicationtoxpowern})$,
\bea
_{(k)} D^{x}_{F} \left(P^{(k)}_n \right) =P^{(k)}_{n-1}.
\eea
The monomials satisfy Theorem 2.1 in \cite{Kac}, and allows one  to derive the Taylor type expansion for arbitrary polynomials.

\begin{thm}($k-$th Golden Taylor expansion)

The derivative operator $_{(k)} D^{x}_{F}$ is a linear operator on the space of polynomials, and
\bea
P_n^{(k)}(x) \equiv \frac{x^n}{F_{n}^{(k)}!} \equiv \frac{x^n}{F_{1}^{(k)} \cdot F_{2}^{(k)} \ldots F_{n}^{(k)} }  \nonumber
\eea
satisfy the following conditions:\\
(i)\phantom{.}$P_0^{(k)}(0)=1$ \phantom{.}and \phantom{.} $P_n^{(k)}(0)=0$ \phantom{.}for \phantom{.}any\phantom{.} $n \geq 1$;\\
(ii)\phantom{.}$deg (P_n^{(k)})= n$; \\
(iii)\phantom{.}$_{(k)} D^{x}_{F}(P_n^{(k)} (x))=P_{n-1}^{(k)} (x)$ for any $n \geq 1$, and $_{(k)} D^{x}_{F}(1)=0.$\\
Then, for any polynomial f(x) of degree N, one has the following Taylor formula;
\bea
f(x)=\sum_{n=0}^N (_{(k)} D^{x}_{F})^{n} f\phantom{.}(0) P_{n}^{(k)} (x)=\sum_{n=0}^N (_{(k)} D^{x}_{F})^{n} f\phantom{.}(0) \frac{x^n}{F_{n}^{(k)} !}. \label{higherordergoldentaylorexpansionformula}
\eea
\end{thm}
As an example
  \bea
\left(x+1\right)^3=24 P_3^{(2)} (x)+9 P_2^{(2)} (x)+3 P_1^{(2)} (x)+P_0^{(2)} (x).
\eea
In the limit $N \rightarrow \infty$ (if it exists) the Taylor formula $(\ref{higherordergoldentaylorexpansionformula})$ determines an expansion of function $ f(x)$ in $P_n^{(k)}(x)$ polynomials,\\
$$f(x)=\sum_{n=0}^\infty (_{(k)} D^{x}_{F}f)^{n}(0) \frac{x^n}{F_{n}^{(k)} !}.$$

\begin{prop}
Let,
\bea
f(z)=\sum_{n=0}^\infty a_n \frac{z^n}{n!} \nonumber
\eea
is an entire complex valued function of complex variable z. Then, for any integer $k$  exists corresponding complex function $_{(k)} \phantom{.} f_{F}(z)$ determined by formula,
\bea
_{(k)} \phantom{.} f_{F}(z)=\sum_{n=0}^\infty a_n \frac{z^n}{F_n^{(k)}!} \nonumber
\eea
and this function is entire.
\end{prop}

\begin{prf}
By the ratio test
\begin{eqnarray}
\rho &=&|z| \lim_{n{\to \infty}} \left|\frac{1}{n+1}\right|  \left|\frac{a_{n+1}}{a_n}\right| = 0 \Rightarrow\\
_{(k)} \rho_{F} &=&|z| \lim_{n{\to \infty}} \left|\frac{1}{F_{n+1}^{(k)}}\right|  \left|\frac{a_{n+1}}{a_n}\right| \\
&=&|z| \lim_{n{\to \infty}} \left|\frac{n+1}{F_{n+1}^{(k)}}\right|  \phantom{.} \left(\left| \frac{1}{n+1}\right| \left|\frac{a_{n+1}}{a_n}\right| \right) \nonumber \\
&=&\lim_{n{\to \infty}} \left| \frac{n+1}{F_{n+1}^{(k)}} \right| \phantom{.} \rho  = 0,\nonumber
\end{eqnarray}
since $\displaystyle{\lim_{n{\to \infty}} \left|\frac{n+1}{F_{n+1}^{(k)}}\right|=0}$.
\end{prf}

As an example of entire complex functions, the following hierarchy of the pair of Golden exponential functions  is introduced
\begin{definition}($k-$th Golden exponentials)
\bea
_{(k)} e^{x}_{F} &\equiv& \sum_{n=0}^\infty  \frac{x^n}{F_{n}^{(k)} !}, \label{expfunction}\\
_{(k)} E^{x}_{F} &\equiv& \sum_{n=0}^\infty  (-1)^{k\frac{n(n-1)}{2}} \frac{x^n}{F_{n}^{(k)} !},\label{Expfunction}
\eea
where
\bea
F_{n}^{(k)}!=F_{1}^{(k)} \cdot F_{2}^{(k)} \cdot F_{n}^{(k)} \ldots F_{n}^{(k)} =\frac{F_{k}\cdot F_{2k}\cdot F_{3k}\ldots F_{nk} }{\left(F_{k}\right)^n }.
\eea
\end{definition}

\begin{prop}
The $k-{th}$ Golden derivative of $k-$th Golden exponentials is
\begin{eqnarray}
_{(k)} D^{x}_{F} \left(_{(k)} e^{\lambda x}_{F}\right)&=&\lambda \phantom{..}   _{(k)} e^{\lambda x}_{F},  \\
_{(k)} D^{x}_{F} \left(_{(k)} E^{\lambda x}_{F}\right)&=&\lambda \phantom{..} _{(k)}  E^{(-1)^k \lambda x}_{F}.
\end{eqnarray}

\end{prop}
This two exponential functions are related by formula
\be
_{(k)}E^{x}_{F} = _{(-k)}e^{x}_{F}.
\ee
 The product of two exponentials is represented by series in powers of $k-$th Golden binomials
 \be
_{(k)}E^{x}_{F} \cdot _{(k)}e^{y}_{F}=  \sum_{n=0}^\infty \frac{_{(k)}(x + y)^n_F}{F_n^{(k)}!}  \equiv _{(k)}e^{_{(k)}(x + y)_F}_{F}.     \label{factorization}
\ee
By exponential function (\ref{Expfunction}) it is possible to introduce translation operator
$_{(k)}E^{y _{(k)}D^x_F}_{F}$ generating these binomials  from monomial $x^n$,
\be
_{(k)}E^{y _{(k)}D^x_F}_{F} x^n = _{(k)}(x + y)^n_F. \label{translation}
\ee
Then, applying this operator to an arbitrary analytic function we get an infinite hierarchy of $k-$th Golden functions
\be
_{(k)}E^{y _{(k)}D^x_F}_{F} f(x) = _{(k)}E^{y _{(k)}D^x_F}_{F} \sum_{n=0}^\infty a_n x^n = \sum_{n=0}^\infty a_n \cdot{_{(k)}(x + y)^n_F}.
\ee
In particular, translated by this operator exponential function (\ref{expfunction}) gives equation (\ref{factorization}),
\be
_{(k)}E^{y _{(k)}D^x_F}_{F} \,\,_{(k)}e^{x}_{F} =  _{(k)}e^{_{(k)}(x + y)_F}_{F},
\ee
or in another form
\be
_{(k)}E^{y _{(k)}D^x_F}_{F} \,\,_{(k)}e^{x}_{F} =   _{(k)}E^{x}_{F} \cdot _{(k)}e^{y}_{F}.
\ee

\subsection{Hierarchy of Golden analytic functions}

 By complex version of translation operator (\ref{translation}) we introduce the complex $k-$th Golden analytic binomials

\be
_{(k)}E^{iy _{(k)}D^x_F}_{F} x^n = _{(k)}(x + i y)^n_F, \label{complextranslation}
\ee
and the hierarchy of $k-$th Golden analytic functions
\be
_{(k)}E^{i y _{(k)}D^x_F}_{F} f(x) =  \sum_{n=0}^\infty a_n \cdot{_{(k)}(x +i y)^n_F} \equiv f(_{(k)}(x + i y)_F),
\ee
for every integer $k$ satisfying the $\bar \partial$-equation
\be
\frac{1}{2}(_{(k)}D^x_F + i _{(-k)}D^y_F)f(_{(k)}(x + i y)_F) = 0.
\ee
The Golden analytic functions from \cite{eski} correspond to particular value $k=1$ of this infinite hierarchy.
The real and imaginary parts of these functions
\be
u(x,y) = _{(-k)} \cos_F( y _{(k)}D^x_F) f(x),\,\,\,\,\,v(x,y) = _{(-k)} \sin_F( y _{(k)}D^x_F) f(x),
\ee
are subject to Cauchy-Riemann equations
\be
_{(k)}D^x_F u(x,y) = _{(-k)}D^y_F v(x,y),\,\,\,\,_{(-k)}D^y_F u(x,y) = - _{(k)}D^x_F v(x,y),
\ee
and are solutions of the hierarchy of Golden Laplace equations

\be
(_{(k)}D^x_F)^2 \phi(x,y) + (_{(-k)}D^y_F)^2 \phi(x,y) = 0.
\ee
Therefore, it  is natural to call these functions as the $k-$th Golden harmonic functions.

\section{Hierarchy of Golden quantum oscillators}
In this section we apply the quantum calculus of Fibonacci divisors or $k-th$ Golden calculus to deformed quantum oscillator problem.
We define the set of creation and annihilation operators $b_k$ and $b^+_k$ in the Fock basis $\{ |n\ra\}$, $n = 0,1,2...$,
represented by infinite matrices
\begin{eqnarray}
 b_k = \left( \begin{array}{cccc} 0  & \sqrt{F^{(k)}_1} & 0 & ...  \\
                                0 & 0 & \sqrt{F^{(k)}_2} & 0     \\
                                0 & 0 &  0 & \sqrt{F^{(k)}_3} \\
                                ... & ... & ... & ... \end{array}\right) ,\,\, b^+_k = \left( \begin{array}{cccc} 0  & 0 & 0 & ...  \\
                                \sqrt{F^{(k)}_1} & 0 & 0 & ...     \\
                                0 & \sqrt{F^{(k)}_2} &  0 & ... \\
                                ... & ... & ... & ... \end{array}\right)\nonumber \label{feq1}
\end{eqnarray}
By introducing the  Fibonacci divisor number operator
\begin{equation}
F^{(k)}_N = \frac{(\varphi^k)^N - (\varphi'^k)^N}{\varphi^k - \varphi'^k},\label{FNk}
\end{equation}
 where $N = a^+ a$ is the standard number operator,  we find that in the Fock basis the eigenvalues of this operator are just Fibonacci divisor numbers conjugate to $F_k$,
\begin{equation}
F^{(k)}_N |n;k \ra= F^{(k)}_n  |n;k \ra \label{Fibnumbereigenvalue}
\end{equation}
and in the matrix form
\begin{eqnarray}
 F^{(k)}_N = \left( \begin{array}{cccc} F^{(k)}_0  & 0 & 0 & ...  \\
                                0 & F^{(k)}_1 & 0 & 0     \\
                                0 & 0 &  F^{(k)}_2 & 0 \\
                                ... & ... & ... & ... \end{array}\right) ,\,\,  F^{(k)}_{N+I} = \left( \begin{array}{cccc} F^{(k)}_1  & 0 & 0 & ...  \\
                                0 & F^{(k)}_2 & 0 & 0     \\
                                0 & 0 &  F^{(k)}_3 & 0 \\
                                ... & ... & ... & ... \end{array}\right). \label{fib1}
\end{eqnarray}
The operator  satisfies recursion rule
$$F^{(k)}_{N+I} = L_k  F^{(k)}_{N} + (-1)^{k-1} F^{(k)}_{N-I} $$
and is expressible as
\begin{equation}
b_k b^+_k = F^{(k)}_{N + I},\,\,\,\,\,\,b^+_k b_k = F^{(k)}_{N},
\end{equation}
giving the commutator
\begin{equation}
[b_k, b^+_k] = F^{(k)}_{N+I} - F^{(k)}_{N}.
\end{equation}

From definition (\ref{FNk}) for  $F^{(k)}_N$ follows matrix identity
\begin{equation}
(\varphi^k)^N = \varphi^k F^{(k)}_N + (-1)^{k+1}F^{(k)}_{N-I},
\end{equation}
where
\begin{eqnarray}
 (\varphi^k)^N = \left( \begin{array}{cccc} 1  & 0 & 0 & ...  \\
                                0 & \varphi^k & 0 & 0     \\
                                0 & 0 &  \varphi^{2k} & 0 \\
                                ... & ... & ... & ... \end{array}\right),\end{eqnarray}
and deformed commutation relations
 \begin{equation}
 b_k b^+_k - \varphi^k b^+_k b_k = \varphi'^{k N}, \,\,\,\,\,\,\,\,\,    b_k b_k^+ - {\varphi'}^k b_k^+ b_k = \varphi^{k N}.
 \end{equation}
The hierarchy of Golden deformed bosonic Hamiltonians, defined as
 \begin{equation}
 { H_k} = \frac{\hbar \omega}{2}( b_k b^+_k + b^+_k b_k) = \frac{\hbar \omega}{2} (F^{(k)}_N + F^{(k)}_{N+I})
 \end{equation}
 is diagonal
  and gives the energy spectrum in terms of Fibonacci divisors
 \begin{equation}
 E^{(k)}_n = \frac{\hbar \omega}{2}(F^{(k)}_n + F^{(k)}_{n+1}) .\label{spectrum}
 \end{equation}
 For even $k$ it becomes
 \be
 E^{(k)}_n = \frac{\hbar \omega}{2} \frac{\sinh [(n+ \frac{1}{2}) k \ln \varphi]}{\sinh [\frac{k}{2}\ln \varphi]}.
 \ee
 In the limit $k \rightarrow 0$ this spectrum  gives the one for linear harmonic oscillator
 \be
 E^{(0)}_n = {\hbar \omega} \left(n+ \frac{1}{2}\right)
 \ee
 and for $k \rightarrow \infty$ it is exponentially growing as powers of Golden ratio
 \be
 E^{(k)}_n \approx\frac{\hbar \omega}{2}  \varphi^{n k}.
 \ee
Following \cite{eski} we can derive
the "semiclassical expansion" of energy levels in powers of $\ln \varphi < 1$,   giving nonlinear corrections to the harmonic oscillator

\be
 E^{(k)}_n = \hbar \omega \left(n+ \frac{1}{2}\right) + 2 \sum_{s=1}^\infty B_{2s+1}(n+1) \frac{k^{2s} (\ln \varphi)^{2s}}{(2s+1)!},
 \ee
 where $B_m (x)$ are Bernoulli polynomials.

The energy levels (\ref{spectrum}) satisfy the three term recurrence relation
 \begin{equation}
 E^{(k)}_{n+1} = L_k E^{(k)}_{n} + (-1)^{k-1} E^{(k)}_{n-1}, \label{threetermrelation}
 \end{equation}
where $L_k$ are Lucas numbers.
The first few energy levels for different values of $k$ are:
\begin{eqnarray}
k&=&1: E^{(1)}_n = 2\frac{\hbar \omega}{2},\, 3 \frac{\hbar \omega}{2},\, 5 \frac{\hbar \omega}{2}, \,8 \frac{\hbar \omega}{2},...\\
k&=&2: E^{(2)}_n = 2\frac{\hbar \omega}{2},\, 21 \frac{\hbar \omega}{2},\, 29 \frac{\hbar \omega}{2},\, 38 \frac{\hbar \omega}{2},...\\
k&=&3: E^{(3)}_n = 5\frac{\hbar \omega}{2},\, 21 \frac{\hbar \omega}{2},\, 89 \frac{\hbar \omega}{2},\, 377 \frac{\hbar \omega}{2},...\\
k&=&4: E^{(4)}_n = 8\frac{\hbar \omega}{2},\, 55 \frac{\hbar \omega}{2}, \,377 \frac{\hbar \omega}{2},\, 2584 \frac{\hbar \omega}{2},...\\
k&=&5: E^{(5)}_n = 12\frac{\hbar \omega}{2},\, 133 \frac{\hbar \omega}{2},\, 1475 \frac{\hbar \omega}{2},\, 16358 \frac{\hbar \omega}{2},...
\end{eqnarray}
The difference  between levels for odd $k = 2l+1$ is growing as
 \begin{equation}
 \Delta E^{(k)}_n = E_{n+1} - E_{n} = \frac{\hbar \omega}{2} L_k F^{(k)}_{n+1},
 \end{equation}
 while for even $k=2l$ it is
 \begin{equation}
 \Delta E^{(k)}_n =  \frac{\hbar \omega}{2} (L_k F^{(k)}_{n+1} - 2 F^{(k)}_n ).
 \end{equation}
Then, the  relative distance
 \begin{equation}
 \frac{\Delta E^{(k)}_n}{E^{(k)}_n} = \frac{F^{(k)}_{n+2} - F^{(k)}_n }{F^{(k)}_{n+1} + F^{(k)}_n}
 \end{equation}
 for asymptotic states $n \rightarrow \infty$
 is given  by the $k$-th power of Golden ratio
 \begin{equation}
 \lim_{n \rightarrow \infty}  \frac{\Delta E^{(k)}_n}{E^{(k)}_n} = \varphi^k -1.
 \end{equation}

For odd $k$ we define the hierarchy of Golden deformed fermionic oscillators by Hamiltonians
\begin{equation}
 { H_k} = \frac{\hbar \omega}{2}(b^+_k b_k -  b_k b^+_k) = \frac{\hbar \omega}{2} (F^{(k)}_N - F^{(k)}_{N+I}),
 \end{equation}
with integer spectrum
\begin{equation}
 E^{(k)}_n = \frac{\hbar \omega}{2}(F^{(k)}_n - F^{(k)}_{n+1}) .\label{fspectrum}
 \end{equation}
In the limit $k \rightarrow 0$ (though it is not odd number) it becomes the usual fermionic two level system with energy $E^{(0)}_n = \frac{\hbar \omega}{2}$ for even $n$,
and $E^{(0)}_n = -\frac{\hbar \omega}{2}$ for odd $n$. In this form it can be applied for description of the qubit. Then,
the deformed case opens possibility to study modifications of  qubits as units of quantum information, depending on  $k$.

The first few members of fermionic hierarchy are follows. For $k=1$ the spectrum is determined just by  Fibonacci numbers $ E^{(1)}_n =  \frac{\hbar \omega}{2}F_{n-1}     $. For $k=3$ and $k=5$ we have an
infinite number of states with energy:
\bea
k&=&3: E^{(3)}_n = 3\frac{\hbar \omega}{2}, \,13\frac{\hbar \omega}{2},\, 55 \frac{\hbar \omega}{2}, \,233 \frac{\hbar \omega}{2},...\\
k&=&5: E^{(5)}_n = 10\frac{\hbar \omega}{2}, \,111 \frac{\hbar \omega}{2}, \,1231 \frac{\hbar \omega}{2}, \,13652 \frac{\hbar \omega}{2},...
\eea

\subsection{Hierarchy of Golden coherent states}

By transformation
 \begin{equation}
 b_k = a\, \sqrt{\frac{F^{(k)}_N}{N}} = \sqrt{\frac{F^{(k)}_{N+I}}{N+I}}\, a,\,\,\,\,\,\,b_k^+ =\sqrt{\frac{F^{(k)}_{N}}{N}}\, a^+ = a^+ \, \sqrt{\frac{F^{(k)}_{N+I}}{N+I}},
 \end{equation}
 we introduce the states
 \begin{equation}
 |n;k \ra_F = \frac{(b_k^+)^n}{\sqrt{F^{(k)}_n!}}|0; k \ra_F, \label{kFock}
 \end{equation}
 where $b_k |0; k \ra_F = 0$,
 coinciding with the Fock states $\{ |n \ra \}$
 and satisfying
 \begin{equation}
 b^+_k |n; k \ra_F = \sqrt{F^{(k)}_{n+1}}\,|n+1, k \ra_F,\,\,\,\,\,b_k\, |n; k \ra_F = \sqrt{F^{(k)}_{n}}\,|n-1; k \ra_F.
 \end{equation}
The hierarchy of Golden coherent states is defined by  eigenstates of annihilation operator
 \begin{equation}
 b_k \, |\beta_k \ra_F = \beta_k \,|\beta_k \ra_F.
 \end{equation}
Expanding in basis (\ref{kFock}), the
normalized coherent states are found as
  \begin{equation}
  |\beta_k \ra_F = \left( _{(k)}e_F^{|\beta_k|^2} \right)^{-1/2} \sum^\infty_{n=0} \frac{\beta_k^n}{\sqrt{F^{(k)}_n!}}|n; k \ra_F,\label{kCoherent}
  \end{equation}
  were the scalar product of two states is
  \begin{equation}
  _F \la\alpha_k|\beta_k \ra_F = \frac{_{(k)}e^{\bar\alpha \beta}_F}{\left( _{(k)}e^{|\alpha|^2}_F\right)^{1/2} \left(_{(k)} e^{|\beta|^2}_F\right)^{1/2}}.
  \end{equation}
The exponential functions in the last two equations are the $k$-th Golden exponentials, defined in  (\ref{expfunction}).

\subsection{Hierarchy of Golden Fock-Bargman representations}

An arbitrary state from the Hilbert space (\ref{kFock}),
$$|\psi \rangle = \sum^\infty_{n=0} c_n |n; k \rangle_F$$
by  the inner product with coherent state (\ref{kCoherent})
  \begin{equation}
  \langle\beta_k| \psi\rangle = \left( _{(k)}e_F^{|\beta_k|^2}\right)^{-1/2}\sum^\infty_{n=0} c_n \frac{\bar \beta^n_k}{\sqrt{F^{(k)}_n!}} = \left( _{(k)}e_F^{|\beta_k|^2}\right)^{-1/2} \psi_k (\bar \beta)
  \end{equation}
  determines the complex analytic wave function
  \begin{equation}
  \psi_k (z) = \sum^\infty_{n=0} c_n \frac{z^n}{\sqrt{F^{(k)}_n!}}
  \end{equation}
  in the corresponding  $k$-th Golden Fock-Bargman representation.
The operators $b_k$ and $b_k^+$ in this
  representation are given by
  \begin{equation}
  b_k \rightarrow\, _{(k)}D^z_F, \,\,\,\, b_k^+ \rightarrow z,
  \end{equation}
  were  the complex $k$-th Golden derivative $ _{(k)}D^z_F$ was  defined in (\ref{definitionofhigherorderderivative}).
Then, the higher order Fibonacci number operator (\ref{FNk}) is represented as
\begin{equation}
  F^{(k)}_N \rightarrow  z\,_{(k)}D^z_F.
  \end{equation}
This way  a link of our Fibonacci divisors Golden calculus with Fock-Bargman representation for hierarchy of deformed quantum
oscillators is established.

As an example we consider the scale invariant analytic functions.
If function $\psi(z)$ is the scale invariant $\psi_s(\lambda z) = \lambda^s \psi_s(z)$,
then it satisfies equation
\begin{equation}
_{(k)}D^z_F \psi_s(z) = \frac{\psi_s (\varphi^k z)- \psi_s({\varphi'}^k z)}{(\varphi^k  - {\varphi'}^k) z} = \frac{ (\varphi^k )^s - ({\varphi'}^k )^s}{(\varphi^k  - {\varphi'}^k) z}  \psi_s(z)
\end{equation}
or
\begin{equation}
z\, _{(k)}D^z_F  \psi_s(z) = F^{(k)}_s \,\psi_s(z).\label{fdifferenceequation}
\end{equation}

This eigenvalue problem is just the $k$-th Golden Fock-Bargman representation of the Fibonacci divisor operator eigenvalue problem (\ref{Fibnumbereigenvalue}), where
eigenfunctions
\begin{equation}
\psi_s(z) = \frac{z^s}{\sqrt{F^{(k)}_s !}}
\end{equation}
are scale invariant.
Then, the  general solution of (\ref{fdifferenceequation}) is of the form
\begin{equation}
\psi_s (z) = z^s f_k(z),
\end{equation}
where $f_k(z)$ is an arbitrary $k$-th Golden-periodic analytic function $f_k(\varphi^k z) = f_k({\varphi'}^k z)$, which has been introduced in Section 3.1.
It was noted that such self-similar wave function characterizes the quantum fractals \cite{Vitiello},\cite{2circle}.

\section{Acknowledgements} This work was  supported by TUBITAK grant 116F206.

\section{Appendix}

Here we proof Theorem 2.2.4. as equation (\ref{maintheorem}) and Theorem 2.2.5. as equation (\ref{mainalpha}).
 First we prove Theorem 2.2.5.
\begin{eqnarray}
F_{kn+k+\alpha}&=&\frac{1}{\varphi-\varphi'}\left[\varphi^{kn+k+\alpha}-\varphi'^{kn+k+\alpha}\right] \nonumber \\
&=&\frac{1}{\varphi-\varphi'}\left[\varphi^{kn+\alpha}\varphi^k-\varphi'^{kn+\alpha}\varphi'^k \right] \nonumber \\
&=&\frac{1}{\varphi-\varphi'}\left[\varphi^{kn+\alpha}\varphi^k+(-\varphi'^{kn+\alpha}\varphi^k+\varphi'^{kn+\alpha}\varphi^k)-\varphi'^{kn+\alpha}\varphi'^k\right]\nonumber\\
&=&\frac{1}{\varphi-\varphi'}\left[(\varphi^{kn+\alpha}-\varphi'^{kn+\alpha})\varphi^k+\varphi'^{kn+\alpha}\varphi^k-\varphi'^{kn+\alpha}\varphi'^k\right]\nonumber\\
&=&F_{kn+\alpha}\varphi^k+\frac{1}{\varphi-\varphi'}\left[\varphi'^{kn+\alpha}\varphi^k-\varphi'^{kn+\alpha}\varphi'^k\right] \nonumber\\
&=&F_{kn+\alpha}\left[\varphi^k+(-\varphi'^k+\varphi'^k)\right]+\frac{1}{\varphi-\varphi'}\left[\varphi'^{kn+\alpha}\varphi^k-\varphi'^{kn+\alpha}\varphi'^k\right]\nonumber\\
&=&F_{kn+\alpha}(\varphi^k+\varphi'^k)-F_{kn+\alpha}\varphi'^k+\frac{1}{\varphi-\varphi'}\left[\varphi'^{kn+\alpha}\varphi^k-\varphi'^{kn+\alpha}\varphi'^k\right]\nonumber\\
&=&F_{kn+\alpha}L_k+\frac{1}{\varphi-\varphi'}\left[\varphi'^{kn+\alpha}\varphi'^k-\varphi^{kn+\alpha}\varphi'^k+\varphi'^{kn+\alpha}\varphi^k-\varphi'^{kn+\alpha}\varphi'^k\right]\nonumber \\
&=&L_k F_{kn+\alpha}+\frac{1}{\varphi-\varphi'}\left[\varphi'^{kn+\alpha}\varphi^k-\varphi^{kn+\alpha}\varphi'^k\right]\nonumber \\
&=&L_k F_{kn+\alpha}+\frac{\varphi^k\varphi'^k}{\varphi-\varphi'}\left[\varphi'^{kn-k+\alpha}-\varphi^{kn-k+\alpha}\right]\nonumber \\
&=&L_k F_{kn+\alpha}-\frac{\varphi^k\varphi'^k}{\varphi-\varphi'}\left[\varphi^{kn-k+\alpha}-\varphi'^{kn-k+\alpha}\right]\nonumber \\
&=&L_k F_{kn+\alpha}-\frac{(\varphi\varphi')^k}{\varphi-\varphi'}\left[\varphi^{kn-k+\alpha}-\varphi'^{kn-k+\alpha}\right] \;\; \mbox{since} \;\; (\varphi\varphi')^k=(-1)^k,\nonumber \\
&=&L_k F_{kn+\alpha}-(-1)^k\left[\frac{\varphi^{kn-k+\alpha}-\varphi'^{kn-k+\alpha}}{{\varphi-\varphi'}}\right]\nonumber
\end{eqnarray}
\begin{eqnarray}
&=&L_k F_{kn+\alpha}+(-1)^{k+1}F_{kn-k+\alpha}\;\; \mbox{and since}\;\;  (-1)^{k+1}(-1)^{-2}=(-1)^{k-1}, \nonumber\\
&=&L_k F_{kn+\alpha}+(-1)^{k-1}F_{kn-k+\alpha} \nonumber
\end{eqnarray}
Therefore, equation (\ref{mainalpha})
\begin{eqnarray}
F_{k(n+1)+\alpha}=L_k F_{kn+\alpha}+(-1)^{k-1}F_{k(n-1)+\alpha} \nonumber
\end{eqnarray} is obtained.
By choosing $\alpha=0$ and dividing both sides of the equation with $F_{k}$ gives us the desired recursion formula $(\ref{maintheorem})$.


\begin{thebibliography}{99}
\bibitem{golden} Pashaev O K and Nalci S 2012 \textit{J Phys A:Math Theor} \textbf{45} 015303
\bibitem{PV1} Parthasarathy, R. and Viswanathan, K.S. , 1991. A q-analogue of the supersymmetric oscillator and its q-superalgebra. \textit{J. Phys. A: Math. Gen.}, Vol. 24, pp.613-617.

\bibitem{PV2} Parthasarathy, R. and Viswanathan, K.S. , 2002. arxiv:hep-th/0212164v1, Dec 2002.

\bibitem{TL1} Tripodi, L. and Lima, C.L. , 1997. On a q-covariant form of the BCS approximation. \textit{Physics Letters B}, Vol. 412, p.7-13.
\bibitem{TL2} Timoteo, V.S. and Lima, C.L , 2006. Effective interactions from q-deformed inspired transformations.  \textit{Physics Letters B},  Vol. 635, p.168-171.

\bibitem{sviratcheva} Sviratcheva, K.D.,  Bahri, C.,  Georgieva, A.I. and Draayer, J.P. , 2004. Physical significance of q-deformation and many-body interactions in nuclei.  \textit{Phys. Rev. Lett.}, Vol. 93, 152501.
\bibitem{ballesteros} Ballesteros, A., Civitarese, O.,  Herranz, F.J. and Reboiro, M. , 2002. Fermion-boson interactions and quantum algebras.  \textit{Phys. Rev. C}, Vol. 66, 064317.
\bibitem{Chaichian} Chaichian, M., Gonzalez Felipe R. and Montonen, C. , 1993. Statistics of q-oscillators, quons and relations to fractional statistics. \textit{J. Phys. A: Math. Gen.}, Vol. 26, pp.4017-4034.

\bibitem{Narayana} Narayana Swamy, P. , 2005. q-deformed Fermions. arxiv:quant-ph/0509136v2, Sept 2005.

\bibitem{Pashaev2019} Pashaev O K 2019 \textit{J. Phys: Conf. Series} \textbf{1194} 012087






\bibitem{2circle} Pashaev O K 2014 \textit{J. Phys: Conf. Series} \textbf{482} 012033
\bibitem{variation} Pashaev O K 2015 \textit{Physica Scripta} \textbf{90} 074010



\bibitem{eski} Pashaev O K 2016 \textit{J. Phys: Conf. Series} \textbf{766} 012015

\bibitem{PashaevYilmaz} Pashaev, O.K. and Yilmaz, O.  , 2008. \newblock Vortex images and q-elementary functions,
\newblock  J.Phys.A:Math.Theor.41, 135207.


\bibitem{PGNJP} Pashaev, O.K. and Gurkan, Z. N. Energy localization  in maximally entangled two- and three-qubit phase space 2012 \textit{New Journal of Physics} \textbf{14} 063007

\bibitem{Vitiello} Vitiello G 2012 \textit{J Physics: Conference Series} \textbf{380} 012021

\bibitem{Faddeev} Faddeev, L.D., Reshetikhin N.Yu. and Takhtajan, L.A. , 1990. Quantization of Lie groups and Lie algebras \textit{Leningrad Math. J.}, Vol. 1, p.193[Alg. Anal.1(1989)187].

\bibitem{Jones} Jones, V.F.R., 1987. Hecke algebra representations of braid groups and link polynomials \textit{Annals of Mathematics}, Vol. 126, p.335.
\bibitem{Kac} Kac, V. and Cheung, P.,2002. Quantum Calculus, Springer, New York.




\bibitem{Arik} Arik, M., Demircan, E., Turgut, T., Ekinci, L. and Mungan, M., 1992. Fibonacci Oscillators.
\textit{Z. Phys. C - Particles and Fields} , Vol. 55, pp.89-95.

\bibitem{Chakrabarti} Chakrabarti, R. and Jagannathan, R., 1991. A (p,q)-oscillator realization of two-parameter quantum algebras. \textit{J. Phys. A: Math. Gen.}, Vol. 24, p.L711.










\end{thebibliography}
\end{document}